\documentclass[12pt]{iopart}

\usepackage{epsfig}
\usepackage{iopams}

\newcommand{\text}[1]{\textrm{#1}}
\newcommand{\bm}{\boldsymbol}
\newcommand{\eqref}[1]{(\ref{#1})}
\newenvironment{subequations}{\numparts}{\endnumparts}

\newcommand{\bo}[1]{\mathbf{#1}}
\newcommand{\Ga}{{\mathcal G}}
\newcommand{\order}{\Or\,}
\newcommand{\mc}[1]{\mathcal{#1}}
\newcommand{\wt}[1]{\widetilde{#1}}
\newcommand{\ba}[1]{\overline{#1}}
\newcommand{\op}[1]{\widehat{#1}}
\newcommand{\dagop}[1]{\widehat{#1}^{\dagger}}
\newcommand{\var}[1]{\text{var}\left[{#1}\right]}
\newcommand{\re}[1]{\text{Re}\left[{#1}\right]}
\newcommand{\im}[1]{\text{Im}\left[{#1}\right]}
\renewcommand{\tr}[1]{\text{Tr}\left[{#1}\right]}

\begin{document}

\title[First-principles quantum dynamics in interacting Bose gases II]
{First-principles quantum dynamics in interacting Bose gases II: Stochastic
gauges }

\author{P Deuar and P D Drummond\footnote{www.physics.uq.edu.au/BEC}}

\address{Australian Centre for Quantum Atom Optics,
    The University of Queensland, Brisbane, Australia}

\eads{\mailto{deuar@physics.uq.edu.au}, \mailto{drummond@physics.uq.edu.au}}

\begin{abstract}
First principles simulations of the quantum dynamics of
interacting Bose gases using the stochastic gauge representation are analyzed.
In a companion paper, we showed how the positive P representation can
be applied to these problems using stochastic differential equations.
That method, however, is limited by increased sampling error as time
evolves. Here, we show how the sampling error can be greatly reduced
and the simulation time significantly extended using stochastic gauges.
In particular, local stochastic gauges (a subset) are investigated.
Improvements are confirmed in numerical calculations of single-, double-,
and multi-mode systems in the weak mode coupling regime. 
Convergence issues are investigated, including the recognition of two 
modes by which stochastic equations produced by phase space methods in general can diverge: 
movable singularities and a noise-weight relationship.
The example calculated here displays wave-like behaviour in spatial correlation functions 
propagating in a uniform 1D gas after a sudden change in the coupling constant.
This could in principle be tested experimentally
using Feshbach resonance methods.
\end{abstract}

\submitto{\JPA}

\section{Introduction}

\label{INTRO}

Phase-space techniques\cite{Wig-Wigner,Hus-Q,Gla-P} for the simulation
of exact many-body quantum dynamics have been developing rapidly\cite{positiveP1,positiveP2,removalofboundaryterms,carusottodynamix,stochasticgauges}. 
Perhaps the major driving force behind this is their ability to work around the exponential growth\footnote{Exponential in the number of modes or orbitals involved.} of Hilbert space that occurs in direct methods that work with an explicit quantum state\cite{Feynman}. This exponential scaling in the direct approach implies that it is virtually impossible to diagonalize general initial states in order to calculate dynamics, even in the exceptional case when the exact eigenstates are known.
Path integral methods were developed to deal with the Hilbert space growth, but while they are very useful for calculating ground states\cite{wilson:74,ceperleypimc}, they are not applicable to quantum dynamics due to the oscillatory phase terms which arise and swamp any observable predictions.
In contrast, appropriate phase-space methods can scale linearly or polynomially with the system size (like path integrals) without being as affected by the oscillatory phases, and so can provide verifiable physical predictions.

 In the companion paper\cite{paperA}, we analyzed
the performance of the nonclassical positive-P representation\cite{positiveP1,positiveP2}
for many-body interacting bosonic systems. Here, we extend the analysis
to treat the more flexible stochastic gauge method of weighted phase-space trajectories\cite{stochasticgauges}.

The aforementioned positive P representation is a straightforward and successful method, 
which is a good ``baseline'' on which more involved methods can build. It is useful to 
give a brief overview of it and its limitations, at this point:  The quantum state is written as
a probability distribution of off-diagonal coherent states separable
at each spatial lattice point. This corresponds to a nonclassical
phase-space with twice the usual classical dimension. The number of
variables needed to specify a single coherent state configuration
grows only \textit{linearly} with the lattice size because they are
separable, while correlations between subsystems are contained in
the details of the distribution (which is stochastically sampled).
This very mild growth of the number of required variables is the reason
that this method can lead to tractable many-body simulations. 

Dynamics in the positive-P representation takes on the particularly simple form
of the widely used Gross-Pitaevskii mean-field equations\cite{GP}, but with appropriate Gaussian
noise terms added. Several interacting Bose gas dynamics simulations with this
method have made quantitative predictions, including:
1) The scattering dynamics during the collision of two BECs, and the evolution of momentum correlations between the scattered atoms\cite{inprep}.
2) Dynamics
of evaporative cooling and incipient condensation of an interacting
Bose gas\cite{evaporativecooling,samperrorinbec}; 
3) Dynamical behaviour of spatial correlations in one-dimensional
interacting Bose gases\cite{paperA,inprep}; 
4) Dynamics of quantum soliton propagation in
optical fibers\cite{carterdrummond:87,Hardman,Solexp}; and
5) Dynamics and
steady-state behaviour of open boson systems coupled to reservoirs\cite{removalofboundaryterms};
Dynamics and thermodynamic spatial correlations of a similar but explicitly
particle-conserving model have also been simulated using the related
stochastic wavefunction method\cite{carusottodynamix,carusottothermo}. 

While much can be simulated with the positive P, there is a limiting factor --- the growth of 
the scatter of trajectories with time. This statistical error can be estimated, but eventually 
after some evolution time it becomes large enough to obscure the physical behaviour for any practical number of trajectories.
Estimates of the useful simulation time with the positive P method
have been obtained\cite{paperA}, and indicate two major limitations: 
\begin{enumerate}
\item When occupation of individual lattice points is high ($\gg\!1$), the
simulation loses precision before phase-collapse can occur.
\item Useful simulation times decrease with reduced lattice spacing. 
\end{enumerate}
This loss of precision is often (but not always) associated with the development of
power-law tails in the distribution function, which can also lead to systematic boundary term errors. A typical cause of power-law tails is the presence of movable singularities in the dimension-doubled phase-space equations (discussed in Section~\ref{BT}). These problems are intimately related, and therefore the methods that improve
precision can also solve boundary term issues.

One approach to this problem is to change the basis. For example, a different phase-space technique that can have lower sampling error when mode occupations are high is the truncated Wigner
representation\cite{Wig-Wigner,Hardman,Steel98}, sometimes called the classical field method\cite{Norrie}. The simulation can be quite accurate, particularly when all modes are strongly occupied. However, this method involves truncation of third-order
differential terms in the evolution. This can lead to large systematic errors which are
essentially intrinsic to the method itself, and tend to grow with time~\cite{Kinsler}.  The conditions under which it is valid for interacting Bose gases have been investigated in significant detail in Ref.~\cite{Sinatra}.

Here, however, we will investigate a different avenue of improvement. For a given basis, there is a wide range of freedom in the correspondence between quantum mechanics and the stochastic equations\cite{stochasticgauges,carusottodynamix}. The available freedoms can be constructively specified in a unified way by the stochastic gauge formalism\cite{stochasticgauges,inprep},
which adds freely defined gauge functions and a corresponding global
weight to the set of stochastic variables.
Preliminary attempts at harnessing this freedom
in single-mode systems have been successful\cite{noiseoptimisation,stochasticgauges},
with demonstrations of time-reversible quantum simulations with up to $10^{23}$
interacting particles\cite{Timereversal}. In this paper we investigate these methods systematically, and also present the progress that has been achieved in exploiting stochastic gauges for large multi-mode systems.

We restrict ourselves to the case of gauge functions defined locally (i.e. separately for each lattice point), which is the simplest useful case.
Sections~\ref{DIFF} and~\ref{DRIFT} investigate two gauge approaches,
each with its own merits. Subsequently, in Sections~\ref{1MODE},~\ref{2MODE},
and~\ref{MMODE}, the performance of the proposed improved methods
is assessed for single-mode, double-mode, and multi-mode cases, respectively.
Section~\ref{BT} investigates convergence issues and identifies two 
ways by which stochastic equations produced by phase space methods in general can diverge: 
movable singularities and a noise-weight relationship.
 The last Section simulates the uniform 1D gas as a non-trivial example.
In particular, the dynamics of second order spatial correlations $\ba{g}^{(2)}(x)$
after a sudden increase in the coupling constant at $t=0$ are calculated as in the companion paper\cite{paperA}.

\section{Model}

\label{METHOD}

\label{SYSTEM}

\subsection{The lattice model}

\label{LATTICEMODEL} We wish to solve for the quantum time-evolution
of a dilute interacting Bose gas. Following the companion paper\cite{paperA},
we consider a lattice Hamiltonian that contains all the essential
features of a continuum model, provided the lattice spacing is sufficiently
small. For a rarefied gas of the kind occurring in contemporary BEC
experiments, $s$-wave scattering dominates\cite{interactionassumptions},
and the $s$-wave scattering length $a_{s}$ is much smaller than
all other relevant length scales. If the lattice spacing is also much
larger than $a_{s}$, then the two-body scattering is well described
by an interaction local at each lattice point. Otherwise, a 
 more careful renormalization procedure\cite{KokkelmansHolland2002} than the one below
is required.

Let us label the spatial dimensions by $d=1,\dots,\mc{D}$, and label
lattice points by the vectors $\bo{n}=(n_{1},\dots,n_{D})$. For lattice
spacings $\Delta{\textrm{x}}_{d}$, the spatial coordinates of the
lattice points are $\bo{x}_{\bo{n}}=(n_{1}\Delta{\textrm{x}}_{1},\dots,n_{D}\Delta{\textrm{x}}_{D})$.
The volume per lattice point is $\Delta V=\prod_{d}\Delta{\textrm{x}}_{d}$.
This lattice implies effective momentum
cutoffs\cite{Abrikosov} of $k^{\rm max}_d=\pi/\Delta x_d$.
We also define the lattice annihilation operators $\op{a}_{\bo{n}}$ ($\approx\sqrt{\Delta V}\,\op{\Psi}(\bo{x}_{\bo{n}})$ in the field notation of Eqn.~(1) of \cite{paperA}),
which obey the usual boson commutation relations of $[\op{a}_{\bo{n}},\dagop{a}_{\bo{m}}]=\delta_{\bo{n}\bo{m}}$.
With these definitions, one obtains: \begin{equation}
\op{H}=\hbar\left[\sum_{\bo{n}\bo{m}}\omega_{\bo{n}\bo{m}}\,\dagop{a}_{\bo{n}}\op{a}_{\bo{m}}+\frac{\kappa}{2}\sum_{\bo{n}}\op{a}^{\dagger2}\op{a}^{2}\right]\,.\label{Hamiltonian}\end{equation}
 In this Hamiltonian, the frequency terms $\omega_{\bo{n}\bo{m}}=\omega_{\bo{m}\bo{n}}^{*}$
come from the kinetic energy and external potential. They produce
a local particle-number dependent energy, and linear coupling to other
sites, the latter arising only from the kinetic processes. The nonlinearity
due to local particle-particle interactions is of strength \begin{equation}
\kappa=\frac{g}{\hbar\Delta V},\end{equation}
 with the standard coupling value\cite{interactionassumptions} being
$g=g_{\textrm{3D}}=4\pi a_{s}\hbar^{2}/m$ in 3D, and $g=g_{\textrm{3D}}/\sigma$
in 2D and 1D, where $\sigma$ is the effective thickness or cross-section
of the collapsed dimensions.

When interaction with the environment is Markovian (i.e. no feedback),
the evolution of the density matrix $\op{\rho}$ can be written as
a master equation\cite{Louisell,gardiner} in Lindblad\cite{Lindblad}
form \begin{equation}
\frac{\partial\op{\rho}}{\partial t}=\frac{1}{i\hbar}\left[\op{H},\op{\rho}\right]+\frac{1}{2}\sum_{j}\left[2\op{L}_{j}\op{\rho}\dagop{L}_{j}-\dagop{L}_{j}\op{L}_{j}\op{\rho}-\op{\rho}\dagop{L}_{j}\op{L}_{j}\right].\label{master}\end{equation}
 For example, single-particle losses at rate $\gamma_{\bo{n}}$ (at
$\bo{x}_{\bo{n}}$) to a \mbox{$T=0$} heat bath are described by
$\op{L}_{\bo{n}}=\op{a}_{\bo{n}}\sqrt{\gamma_{\bo{n}}}$.

\subsection{Gauge P representation}

\label{GAUGEP} The gauge P representation was introduced in Ref.~\cite{removalofboundaryterms}
and is described in more detail in Refs.~\cite{stochasticgauges,inprep}.
Here we summarize the issues relevant to the dynamics of the model
(\ref{Hamiltonian}), and present the stochastic equations to simulate. 

For a lattice with $M$ points, the density matrix is expanded as
\begin{equation}
\op{\rho}=\int G(\bm\alpha,\bm\beta,\Omega)\,\op{\Lambda}(\bm\alpha,\bm\beta,\Omega)\, d^{2M}\!\bm\alpha\, d^{2M}\!\bm\beta\, d^{2}\Omega,\label{gaugerep}\end{equation}
 where $\op{\Lambda}$ is an off-diagonal operator kernel, separable
between the $M$ lattice point subsystems. We use a coherent state
basis so that \begin{equation}
\op{\Lambda}(\bm\alpha,\bm\beta,\Omega)=\Omega\,||\,\bm\alpha\rangle\langle\bm\beta^{*}||\exp[-\bm\alpha\cdot\bm\beta],\label{kernel}\end{equation}
 in terms of Bargmann coherent states with complex amplitudes $\bm\alpha=(\dots,\alpha_{\bo{n}},\dots)$:
\begin{equation}
||\,\bm\alpha\rangle=\otimes_{\bo{n}}\exp\left[\alpha_{\bo{n}}\dagop{a}_{\bo{n}}\right]|\,0\rangle.\end{equation}
 $\Omega$ is a complex weight. 

The gauge representation is a generalization of the positive P representation,
which has no weight term. With the choice \begin{equation}
G(\bm\alpha,\bm\beta,\Omega)=P_{+}(\bm\alpha,\bm\beta)\,\delta^{2}(\Omega-1),\label{posPtoG}\end{equation}
 we recover the positive P distribution $P_{+}$. It has been shown
that all density matrices can be represented by a positive real $P_{+}$\cite{positiveP2},
so the same is true for $G$. The expansion (\ref{gaugerep}) then
becomes a probability distribution of the $\op{\Lambda}$, or equivalently
the variables $\bm\alpha$, $\bm\beta$, $\Omega$. A constructive
expression for a distribution $P_{+}(\op{\rho})$ is given by expression
(3.7) in Ref.~\cite{positiveP2}, although more compact distributions
may exist. In particular, a coherent state $|\bm\alpha_{o}\rangle$
will simply have \begin{equation}
G=\delta^{2M}(\bm\alpha-\bm\alpha_{o})\,\delta^{2M}(\bm\beta-\bm\alpha_{o}^{*})\,\delta^{2}(\Omega-1).\end{equation}

Using the identities \begin{subequations}\label{id}\begin{eqnarray}
\op{a}_{\bo{n}}\op{\Lambda} & = & \alpha_{\bo{n}}\op{\Lambda},\label{ida}\\
\dagop{a}_{\bo{n}}\op{\Lambda} & = & \left[\beta_{\bo{n}}+\frac{\partial}{\partial\alpha_{\bo{n}}}\right]\op{\Lambda},\label{idad}\\
0 & = & \left[1-\Omega\frac{\partial}{\partial\Omega}\right]\op{\Lambda},,\label{ido}\end{eqnarray}
\end{subequations}
 the master equation (\ref{master}) in $\op{\rho}$ can be shown
to be equivalent to a Fokker Planck equation in $G$, and then to
stochastic equations in the $\alpha_{\bo{n}}$, $\bm\beta_{\bo{n}}$,
and $\Omega$, provided that boundary terms vanish in carrying
out the partial integration used to
derive the Fokker-Planck equation. This implies a restriction on the
phase-space distribution tails, which typically must vanish faster than any power law.

The standard method is described in Ref.~\cite{gardiner}.
The correspondence is in the sense that appropriate stochastic averages
of these variables correspond to quantum expectation values in the
limit when the number of trajectories $\mc{S}\rightarrow\infty$.
In particular one finds that\cite{removalofboundaryterms}\begin{equation}
\left\langle \prod_{jk}\dagop{a}_{\bo{n}_{j}}\op{a}_{\bo{n}_{k}}\right\rangle =\lim_{\mc{S}\rightarrow\infty}\frac{\left\langle \prod_{jk}\beta_{\bo{n}_{j}}\alpha_{\bo{n}_{k}}\Omega+\prod_{jk}\alpha_{\bo{n}_{j}}^{*}\beta_{\bo{n}_{k}}^{*}\Omega^{*}\right\rangle _{\textrm{s}}}{\left\langle \Omega+\Omega^{*}\right\rangle _{\textrm{s}}}\,\,.\label{moments}\end{equation}
 Any observable can be written as a linear combination of terms \eqref{moments}.
We will use the notation $\langle\cdot\rangle_{\textrm{s}}$ to distinguish
averages of random variables from quantum expectations $\langle\cdot\rangle$. 

Due to the fact that the basis vectors $||\bm\alpha\rangle$ are not
mutually orthogonal, many different distributions (and hence, sets
of equations) correspond to the same master equation. It has been
found\cite{stochasticgauges,inprep} that the family of stochastic
equations corresponding to a given master equation is 
parameterized by \textit{stochastic gauge functions} of several kinds.
These are completely arbitrary functions that appear in the equations,
but do not affect (\ref{moments}). However, the rate at which precision
of the estimates (\ref{moments}) improves with more stochastic realizations
can be strongly affected. It is shown below that judicious choices
of the gauges can lead to large improvements in precision. 

The $2M+1$ Ito stochastic equations to simulate are found using methods
described in Refs.~\cite{gardiner,stochasticgauges}. For the model
(\ref{Hamiltonian}) obeying the master equation (\ref{master}) with
coupling to a zero temperature heat bath, they are \begin{subequations}\label{itoequations}\begin{eqnarray}\fl
d\alpha_{\bo{n}}  = & [-i\sum_{\bo{m}}\omega_{\bo{n}\bo{m}}\alpha_{\bo{m}}-i\kappa n_{\bo{n}}\alpha_{\bo{n}}-\gamma_{\bo{n}}\alpha_{\bo{n}}/2]dt+\sum_{k}B_{\bo{n}k}^{(\alpha)}(dW_{k}-\Ga_{k}dt), \\\fl
d\beta_{\bo{n}}  = & [i\sum_{\bo{m}}\omega_{\bo{n}\bo{m}}^{*}\beta_{\bo{m}}+i\kappa n_{\bo{n}}\beta_{\bo{n}}-\gamma_{\bo{n}}\beta_{\bo{n}}/2]dt+\sum_{k}B_{\bo{n}k}^{(\beta)}(dW_{k}-\Ga_{k}dt),\\\fl
d\Omega  = & \Omega\left[\sum_{k}\Ga_{k}dW_{k}\right].\label{itodOmega}\end{eqnarray}
\end{subequations}
 Here, $n_{\bo{n}}=\alpha_{\bo{n}}\beta_{\bo{n}}$, and there are
$M'\geq2M$ labels $k$ for the independent noise terms, to sum over.
The $dW_{k}$ are independent Wiener increments $\langle dW_{j}(t)\, dW_{k}(s)\rangle_{\textrm{s}}=\delta_{jk}\delta(t-s)dt^{2}$.
In practice, these can be realized at each time step $\Delta t$ by
mutually- and time-independent real Gaussian noises of mean zero,
and of variance $\Delta t$. The elements of the $M\times M'$ \textit{noise
matrices} $B$ must satisfy \begin{subequations}\label{BB=3DD}\begin{eqnarray}
\sum_{k}B_{\bo{n}k}^{(\alpha)}B_{\bo{m}k}^{(\alpha)} & = & -i\kappa\alpha_{\bo{n}}^{2}\delta_{\bo{n}\bo{m}}\label{BB=D1}\\
\sum_{k}B_{\bo{n}k}^{(\beta)}B_{\bo{m}k}^{(\beta)} & = & i\kappa\beta_{\bo{n}}^{2}\delta_{\bo{n}\bo{m}},\label{BB=D2}\\
\sum_{k}B_{\bo{n}k}^{(\alpha)}B_{\bo{m}k}^{(\beta)} & = & 0.\end{eqnarray}\end{subequations}

The $M$ complex quantities $\Ga_{k}$ are \textit{arbitrary} complex
gauge functions, referred to as \textit{drift stochastic gauges}.
There is also more gauge freedom here because (\ref{BB=3DD}) do not
specify the noise matrices $B^{(\alpha)}$ and $B^{(\beta)}$ precisely.
This freedom can be expressed as \textit{diffusion stochastic gauges}\cite{noiseoptimisation,removalofboundaryterms,stochasticgauges}. 

The simulation strategy is (briefly):
\begin{enumerate}
\item Sample a trajectory according to the known initial condition $G(0)=G(\,\op{\rho}(0)\,)$
\item Evolve according to the stochastic equations (\ref{itoequations}),
calculating moments of interest, and recording. 
\item Repeat for $\mc{S}\gg1$ independent trajectories and average. 
\end{enumerate}

\subsection{Single-mode model}

\label{SINGMODE} The single mode model is a good test bed for gauge
choices in quantum many-body theory. It is exactly solvable, yet it
already contains all the essential features that lead to the rapid
growth of fluctuations which limits simulation time with the positive
P method\cite{paperA}. It has also been experimentally realized in
recent optical lattice experiments\cite{Greiner} with Bose-Einstein
condensates. We proceed as in the companion paper\cite{paperA}:

To simplify the notation, the mode labels $\bo{n}=(0,\dots,0)$ will
be omitted when referring to the single mode system. Furthermore,
we move to an interaction picture where the harmonic oscillator evolution
due to the $\omega\dagop{a}\op{a}$ term in the Hamiltonian is implicitly
contained within the Heisenberg evolution of the operators. Then,
this {}``anharmonic oscillator'' model simply has \begin{equation}
\op{H}=\frac{\hbar\kappa}{2}\op{a}^{\dagger2}\op{a}^{2}.\label{1modeH}\end{equation}

When dealing with this system it will be convenient to consider the
evolution starting with an off-diagonal coherent-state kernel \begin{equation}
\op{\Lambda}_{0}=\op{\Lambda}(\alpha_{0},\beta_{0},1),\label{offkernel}\end{equation}
 with {}``particle number'' $n_{0}=\alpha_{0}\beta_{0}$ and initial
unit weight $\Omega(0)=1$. This is because for any general initial state, each sampled trajectory will start out as an $\op{\Lambda}_0$ with some coherent amplitudes.

With initial condition $\op{\Lambda}_{0}$,
analytic expressions for observables can be readily obtained. In particular,
the first-order time-correlation function \begin{equation}
G^{(1)}(0,t)=\beta_{0}\langle\op{a}\rangle=\alpha_{0}^{*}\langle\dagop{a}\rangle^{*},\label{G1tdef}\end{equation}
 which contains phase coherence information. Normalizing by $\text{Tr}[\op{\Lambda}_{0}]$,
\begin{equation}
G^{(1)}(0,t)=n_{0}\, e^{-\gamma t/2}\,\exp{\left\{ \frac{n_{0}}{1-i\gamma/\kappa}\left(e^{-i\kappa t-\gamma t}-1\right)\right\} }.\label{G1exact}\end{equation}

When the damping is negligible, $n_{0}$ real, and the number of particles
is $n_{0}\gg1$, one sees that the initial phase oscillation period
is \begin{equation}
t_{\text{osc}}=\frac{1}{\kappa}\sin^{-1}\left(\frac{2\pi}{n_{0}}\right)\approx\frac{2\pi}{\kappa n_{0}},\end{equation}
 and the phase collapse time\cite{Steel98} over which $|G^{(1)}(0,t)|$
decays\footnote[1]{We have kept the notation $t_{\rm coh}$ from the companion paper\cite{paperA}, where ``coh'' indicates that 
coherence is maintained.} is \begin{equation}
t_{\text{coh}}=\frac{1}{\kappa}\cos^{-1}\left(1-\frac{1}{2n_{0}}\right)\approx\frac{1}{\kappa\sqrt{n_{0}}}.\label{tcoh}\end{equation}
 When there is no damping, the first quantum revival occurs - as observed
experimentally\cite{Greiner} - at \begin{equation}
t_{\text{revival}}=\frac{2\pi}{\kappa}.\end{equation}

\subsection{Positive P noise behaviour}

\label{USEFUL} The noise behaviour of this model and simulation times
has been investigated in detail for the positive P case in Ref.~\cite{paperA}.
Some useful conclusions include: 

\begin{enumerate}
\item If $\xi$ is a Gaussian random variable, and $v=e^{\sigma\xi}$, then
finite-sample estimates of $\langle v\rangle_{\textrm{s}}$ using
$\mc{S}$ realizations of the random variable have relative fluctuations
that scale as $\propto\sqrt{(e^{\sigma^{2}}-1)/\mc{S}}$. Due to
the rapid rise after $\sigma\gtrsim1$, reasonable precision can only
be obtained with practical sample sizes ($\mc{S}\lesssim10^{6}$,
say) while \begin{equation}
\sigma^{2}\lesssim10.\end{equation}

\item In a single mode system, the variable $n(t)$ is the exponential of
a Gaussian random variable, as is $\alpha(t)$ at short times. Since
observable estimates \eqref{moments} involve means of polynomial
functions $f$ of these variables $\langle f\rangle_{\textrm{s}}$,
then one needs \begin{equation}
\var{\log|f|}\lesssim\order(10)\label{varflimit}\end{equation}
 for reasonable precision. 
\item The mechanism responsible for limiting simulation time is growth of
fluctuations in $\log|\alpha|$ due to the real part of the $d\log\alpha=-i\kappa n\, dt$
term. When $|n|\gg1$, the fluctuations in $\im{n}$ are sizeable,
and $\var{\log|\alpha|}$ rapidly reaches large values that exceed
\eqref{varflimit}. 
\item In the multi-mode system, this nonlinear (in $\alpha_{\bo{n}}$) term becomes
$d\alpha_{\bo{n}}=-i\kappa n_{\bo{n}}\alpha_{\bo{n}}\, dt$, which
depends only on the variables in the local mode $\bo{n}$. For this
reason, the single mode analysis continues to give a good qualitative
description of the noise behaviour in multi-mode systems, especially
while the inter-mode coupling is weak in comparison with the local
scattering. 
\end{enumerate}

\subsection{Aims}

\label{AIMS}

In what follows, an estimate $t_{\rm sim}$ of the useful simulation time will be evaluated using the condition (\ref{varflimit}) for the amplitude variables $\alpha$ and $\beta$. 
For the positive P method applied to the system (\ref{1modeH}), $t_{\rm sim}$ 
was found to be relatively shortest at high mode occupation and weak
damping, scaling as $\propto n^{-2/3}$. This is shown in Figure~\ref{FIGtsim}
and Table~\ref{TABLEtsim}. For $n\gg1$ we see that the positive
P method does not reach even the phase collapse times $t_{\rm coh}$\cite{Steel98}.
This is caused by the phase-diffusion inherent in any initial state
with a range of particle numbers, where the self-interactions cause
the phase to evolve at a different rate for different total particle
number. 

However, indications
that simulation improvements can be obtained using nonzero gauge choices were seen
by Deuar and Drummond\cite{ccp2k} and Plimak \textit{et al} \cite{noiseoptimisation},
using drift and diffusion modifications, respectively. Carusotto \textit{et
al}\cite{carusottodynamix} also found improvements in a different
(number-conserving) model, using an approach that effectively uses
a drift gauge (see Section~\ref{OTHERGAUGE}, for more on this). 

The aim here is to develop gauges that improve simulation times at
high mode occupation, while being applicable to the many-mode situation
where conditions for a single mode are dynamically changing. To do
this we wish to find the dependence of advantageous gauges on mode
parameters such as $n(t)$, constants $\kappa$, $\gamma$, and time
$t$ itself. We are especially interested in the case of low or absent
damping $\gamma\ll\kappa$, in highly occupied modes $n\gg1$. This
is the regime where the simulation time is most limited with the positive
P method.

\section{Local diffusion gauges}

\label{DIFF}

In the positive P simulation of modes with $n\gg1$ occupations, $t_{\rm sim}$ is limited by the following process:
In order to allow for an exact simulation
of quantum fluctuations, the nonclassical
phase space requires the effective particle number $n=n'+in''$ to
develop complex values. This in turn gives rise to exponential growth
in amplitude fluctuations in either $|\alpha|$ or $|\beta|$, since
the drift equations have the structure:\begin{subequations}\begin{eqnarray}
\frac{d|\alpha|}{dt} & = & \kappa n''|\alpha| + \dots  \\
\frac{d|\beta|}{dt} & = & -\kappa n''|\beta| + \dots\,\,\label{eq:instability}\end{eqnarray}
\end{subequations}
These  instabilities lead to growth in sampling error, and possibly ultimately
to systematic boundary term errors as well. 
In this section we consider the freedoms present in the choice of
noise matrix $B$, to limit the stochastic growth in the effective
`gain' $n''$, without introducing any drift gauges. Such techniques can extend the
useful time-scale of a simulation, although they are typically unable to remove
boundary terms at long times if caused by movable singularities.

In this case $\mc{G}_{k}=0$,
$d\Omega=0$, so there is no weight evolution and we can assume $\Omega=1$
for all practical purposes. Analytic expressions for an optimized
diffusion gauge choice will be found and discussed. Their performance
in actual numerical simulations is reported in Sections~\ref{1MODE}
to~\ref{MMODE}.

\subsection{Diffusion gauge mechanism}

\label{DIFFUSION} The noise matrices appearing in the stochastic
equations must obey (\ref{BB=3DD}) which can be written as a matrix
equation $BB^{T}=D$, where \begin{equation}
B=\left[\begin{array}{c}
B^{(\alpha)}\\
B^{(\beta)}\end{array}\right],\end{equation}
 and $D$ is completely determined by the master equation. Otherwise
the $B$ are free, including the number of noises $M'$, i.e. columns
in the matrix. These freedoms in $B$ choices are investigated in
Ref.~\cite{inprep}. A broad class of suitable $B$ which are square
complex orthogonal matrices have been described in \cite{stochasticgauges}.
These are given by \begin{equation}
B(g_{jk})=\sqrt{D}\, O(g_{jk}),\end{equation}
 where $O$ are arbitrary complex $2M\times2M$ orthogonal matrices
such that $OO^{T}=I$. In the general case, these can be constructed
using $(2M-1)M$ complex \textit{diffusion gauge} functions $g_{jk}$:
\begin{equation}
O=\exp\left(\sum_{j<k}g_{jk}\sigma^{(jk)}\right),\end{equation}
 where the antisymmetric matrix basis ($j\neq k=1,\dots,2M)$ has
elements \begin{equation}
\sigma_{lp}^{(jk)}=\delta_{jl}\delta_{kp}-\delta_{jp}\delta_{kl}.\end{equation}

\subsection{Single mode}

\label{1MODEGII} We proceed similarly to the companion paper\cite{paperA}
for the positive P case. For the model \eqref{1modeH} with damping
to a zero temperature heat bath, the equations to simulate are 
\begin{eqnarray}\eqalign{\label{1modegiequations}
d\alpha =  -&\alpha[i\kappa n+\gamma/2]\, dt
+i\sqrt{i\kappa}\,\alpha\,[\cos(g)dW_{1}+\sin(g)dW_{2}]\\
d\beta =  &\beta[i\kappa n-\gamma/2]\, dt
+\sqrt{i\kappa}\,\beta\,[-\sin(g)dW_{1}+\cos(g)dW_{2}].}\end{eqnarray}
 with $n=\alpha\beta$. Here we include only one complex diffusion
gauge $g=g_{12}=g'+ig''$.

\subsubsection{Real gauges}

Note that \begin{equation}
B=B_{o}O(g)=B_{o}S(g'')R(g').\end{equation}
 in terms of a rotation $R$ and a transformation $S$. The rotation
serves only to mix the noises $dW_{j}$ together: \begin{equation}
R(g')dW=\left[\begin{array}{cc}
\cos g' & \sin g'\\
-\sin g' & \cos g'\end{array}\right]\left[\begin{array}{c}
dW_{1}\\
dW_{2}\end{array}\right]=\left[\begin{array}{c}
d\wt{W}_{1}\\
d\wt{W}_{2}\end{array}\right].\end{equation}
 where the new noises $d\wt{W}_{j}$ have the same statistical properties
as the original noises $dW_{j}$. Because of this, $g'$ has no impact
on the statistical properties of the simulation. Henceforth we will
consider only $g=ig''$, where $\cos g=\cosh g''$, and $\sin g=i\sinh g''$.

\subsubsection{Logarithmic variance}

As in \cite{paperA} for the standard positive P case, we consider
the mean phase variable variance \begin{equation}
\mc{V}=\frac{1}{2}\left\{\var{\log|\alpha(t)|}+\var{\log|\beta(t)|}\right\},\label{mcVdef}\end{equation}
 which was found to be the limiting factor for simulation time when
it exceeds $\mc{V}\gtrsim\order(10)$ as per \eqref{varflimit}. 

Taking $g=ig''$ the equations \eqref{1modegiequations} can be formally
solved using the rules of Ito calculus as \begin{subequations}\label{formalpp}\begin{eqnarray}
\fl\log n(t) =  \log n(0)-\gamma t+\sqrt{i\kappa}\,e^{-g''}\left[\zeta_{2}(t)+i\zeta_{1}(t)\right],\qquad\label{nformalpp}\\
\fl\log\alpha(t)  =  \log\alpha(0)+(i\kappa-\gamma)t/2+i\sqrt{i\kappa}\,\zeta_{1}(t)\cosh g''
 -\sqrt{i\kappa}\,\zeta_{2}(t)\sinh g''-i\kappa\int_{0}^{t}n(s)\, ds, \nonumber\\&\end{eqnarray}
 where \begin{equation}
\zeta_{j}(t)=\int_{s=0}^{t}dW(s)\label{zetadef}\end{equation}\end{subequations}
 are Gaussian-distributed random variables of mean zero, and: \begin{equation}
\langle\zeta_{j}(t)\zeta_{k}(s)\rangle_{\textrm{s}}=\delta_{jk}\text{min}\left[t,s\right].\label{zetamean}\end{equation}

It is convenient to define \begin{equation}
n_{0}=\alpha_{0}\beta_{0}=n'_{0}+in''_{0}.\end{equation}
 Note that $n(t)$ is the exponential of a Gaussian random variable,
and that if a generic variable $\xi$ is Gaussian then \begin{equation}
\lim_{\mc{S}\rightarrow\infty}\langle\xi(t)^{m}\rangle_{\textrm{s}}=\frac{m!}{2^{m/2}(m/2)!}\label{ximoments}\end{equation}
 for even $m$, and zero for odd $m$. From this, $\lim_{\mc{S}\rightarrow\infty}\langle e^{\sigma\xi}\rangle_{s}=\exp(\sigma^{2}/2)$
and $\lim_{\mc{S}\rightarrow\infty}\langle\cos(\sigma\xi)\rangle_{s}=\exp(-\sigma^{2}/2)$.
Using these we obtain that the limiting variance $\mc{V}$ in the limit of a large
number of samples $\mc{S}\to\infty$ is given by:
\begin{subequations}
\begin{eqnarray}\label{vpp}
\mc{V} & = & \kappa t\cosh(2g'')/2-\kappa^{2}n'_{0}\left\{ \frac{1-e^{-\gamma t}(1+\gamma t)}{\gamma^{2}}\right\} \\
 & + & \frac{\kappa^{2}|n_{0}|^{2}}{q\gamma}\left\{ 1-\frac{qe^{-\gamma t}-\gamma e^{-qt}}{(q-\gamma)}-\frac{q}{2\gamma}\left(1-e^{-\gamma t}\right)^{2}\right\} ,\nonumber \end{eqnarray}
where \begin{equation}
q=2(\gamma-\kappa e^{-2g''})\end{equation}
\end{subequations}
 is a {}``damping strength'' parameter. 

At short times the first term (direct fluctuations from noise in $d\log|\alpha|$ etc.) certainly dominates.
For long time-scales the indirect noise mediated by a growing spread in $n''$ (remaining terms) has the most serious consequences, since it causes an exponential growth in the sampled fluctuations.

\subsubsection{Optimum gauge}

The first term in \eqref{vpp} (due to direct noise in $\log\alpha$
and $\log\beta$) grows with $g''$, while the later terms decrease.
This indicates that there is a trade-off parameterized by $g''$.
The optimum choice is given by a balance between the direct noise
in the amplitudes, and the indirect noise in the gain term $n''$. 
The lowest fluctuations $\mc{V}$ at a given time $t$ will occur
when $\partial\mc{V}(t)/\partial g''=0$. To choose $g''$, we must decide
upon a {}``target time'' $t=t_{\textrm{opt}}$ at which to minimize
$\mc{V}$.

\begin{figure}
\center{\includegraphics[width=250pt]{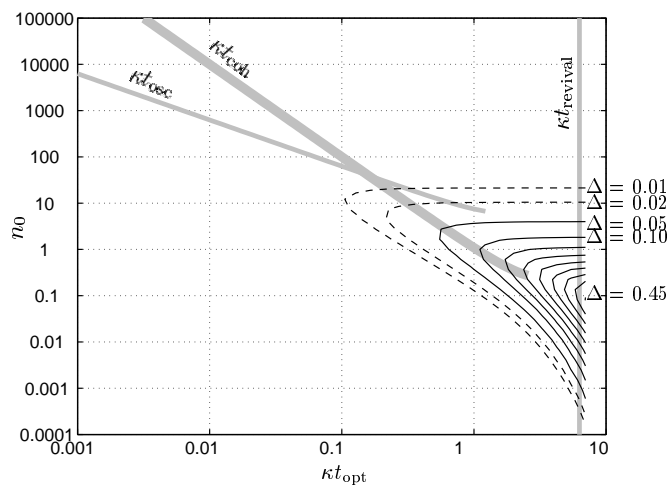}}\vspace*{-10pt}
\caption{\label{FIGgiidisc} Discrepancy $\Delta=g''_{\text{opt}}-g''_{\text{approx}}$
between $g''_{\textrm{opt}}$ (the exact optimization of $g''$ for
the diffusion-gauge-only case) and the approximate expression (36)
for an undamped mode with diagonal coherent initial occupation $n_{0}=n'_{0}$.
Discrepancy values $\Delta$ are shown as contours. }
\end{figure}

When we aim for relatively short target times $t_{\textrm{opt}}$
such that $|q|t_{\textrm{opt}}\ll1$ and $2\kappa t_{\textrm{opt}}e^{-2g''}\ll1$,
the optimized gauge is approximated by \begin{subequations}\label{noptgii}\begin{equation}
g''_{\textrm{opt}}\approx g''_{\textrm{approx}}=\frac{1}{4}\log\left[\frac{(2\kappa t_{\textrm{opt}}|n_{0}|)^{2}}{3}\, a_{2}(\gamma t_{\textrm{opt}})+1\right],\end{equation}
 where, on defining $v=\gamma t_{opt}$, the coefficient $a_{2}(v)$
is \begin{equation}
a_{2}(v)=\frac{3}{2}\left(\frac{e^{-2v}(3+2v)+1-4e^{-v}}{v^{3}}\right).\label{a2def}\end{equation}\end{subequations}
 This reduces to $a_{2}(0)=1$ in the undamped case. The discrepancy
between (36) and the exact optimization $g''_{\textrm{opt}}$
found using $\partial\mc{V}/\partial g''=0$ is shown for real $n_{0}=n'_{0}$
in Figure~\ref{FIGgiidisc}. It can be seen that for $n'_{0}\gtrsim\order(10)$
and/or for times shorter than singly-occupied coherence time $1/\kappa$,
the approximate expression for the optimized gauge choice is still
useful. 

\subsubsection{Useful simulation times}

By inspection of (31), 
the behaviour of $\log|\alpha|$
and $\log|\beta|$ is Gaussian-like due to the $\zeta_{j}$ terms.
 Since
$G^{(1)}\propto\langle\alpha\rangle_{\textrm{s}}$, the condition \eqref{varflimit}
implies that useful precision in $G^{(1)}$ is obtainable only while
$\mc{V}\lesssim10$.

In the simplest case of no damping and coherent state initial conditions at large occupation $n_0=n'_0\gg1$, for target times
times $t_{\rm opt} \gg 3t_{\rm osc}/4\pi$ one has $e^{-2g''_{\rm approx}} \approx \sqrt{3}/2\kappa t_{\rm opt}n_0$. These are times longer than an oscillation period but shorter than coherence time $t_{\rm coh}$.
Taking then the terms of $\mc{V}(t,g'')$ from (36) 
at $t=t_{\rm opt}$ of highest order in $n_0$, and imposing the precision limit $\mc{V}(t_{\rm sim},g''_{\rm approx})\lesssim10$, one obtains the expected simulation time for an undamped
system as \begin{equation}
t_{\textrm{sim}}\approx\order(10)\, t_{\textrm{coh}}.\label{giitsim}\end{equation}
 This is a large improvement at high mode occupations --- compare to the positive P ($g''=0$) results in Table~\ref{TABLEtsim}. 

Checking back, we see that when the target time is given by $t_{\rm opt}=t_{\rm sim}$ from \eqref{giitsim}, it is indeed much longer than an oscillation period, and also that the terms of $\mc{V}$ highest order in $n_0$ remain so after taking $\kappa t\lesssim\,\propto n_0^{-1/2}$ into account.
The clearest evidence for the validity of the above reasoning --- which
involves a few approximations --- is that the numerical simulations
of Section~\ref{1MODE} agree with \eqref{giitsim}. 

At low occupations and with $n_{0}=n'_{0}$, on the other hand, $g''_{\textrm{approx}}\rightarrow0$,
and one again expects the same simulation time as with the standard
positive P equations.

\subsubsection{Strong damping}

If damping is present, then with a large enough gauge \mbox{$g''>\frac{1}{2}\log(\kappa/\gamma)$}
such that $q>0$ the regime of linear increase in sampling variance can always be reached for times $qt\gg1$.
Here, fluctuations grow slowly as \begin{equation}
\mc{V}=\kappa t\cosh(2g'')/2-b\,\,.\end{equation}
The constant $b$ is \begin{equation}
b = \frac{\epsilon^2}{2}\left[2n'_0-|n_0|^2\epsilon/(e^{2g''}-\epsilon)\right],\label{bdefgii}\end{equation}
 where $\epsilon=\kappa/\gamma<e^{2g''}$. The required $q>0$ implies 
$\gamma>\kappa e^{-2g''}$, i.e. either that damping rates are strong compared
to the nonlinear detuning at the two-particle level, or the diffusion gauge is large.

\subsection{Extension to many-modes}

\label{ADAPT} The expression (36) 
 was worked out under
the assumption that the mean occupation of the mode is conserved.
In coupled-mode simulations this is no longer the case, and can be
adapted for by replacing $n_{0}$ in the expression (36) 
by $n_{\bo{n}}(t)$. 

Next, consider the situation when we aim to simulate until a target
time $t_{\textrm{opt}}$. At some intermediate time $0<t<t_{\textrm{opt}}$
it may be advantageous to optimize $g''$ only for the \textit{remaining}
time to target \begin{equation}
t_{\textrm{rem}}=\text{max}[t_{\text{opt}}-t,0],\label{tremdef}\end{equation}
 rather than for a constant target time $t_{\textrm{opt}}$ ahead
of $t$. Indeed, this is demonstrated to be an improvement in Section~\ref{TARGETT}. 

There will be a separate local gauge at each lattice point, such that
for an $M$-mode system (with $j=1,\dots,M$ labeling the modes) the
only nonzero noise matrix elements are \begin{eqnarray}\eqalign{\label{Bdef}
B_{\bo{n}_{j},j}^{(\alpha)}= &i\sqrt{i\kappa}\alpha_{\bo{n}_{j}}  \cosh g''_{\bo{n}_{j}} \\
B_{\bo{n}_{j},(j+M)}^{(\alpha)}= -&\sqrt{i\kappa}\alpha_{\bo{n}_{j}}  \sinh g''_{\bo{n}_{j}} \\
B_{\bo{n}_{j},j}^{(\beta)}=  -&i\sqrt{i\kappa}\beta_{\bo{n}_{j}}  \sinh g''_{\bo{n}_{j}} \\
B_{\bo{n}_{j},(j+M)}^{(\beta)}= &\sqrt{i\kappa}\beta_{\bo{n}_{j}}  \cosh g''_{\bo{n}_{j}}.}\end{eqnarray}
 The suggested diffusion gauges are \begin{equation}
g''_{\bo{n}}=\frac{1}{4}\log\left[\frac{(2\kappa t_{\textrm{rem}}|n_{\bo{n}}(t)|)^{2}}{3}\, a_{2}(\gamma_{\bo{n}}t_{\textrm{rem}})+1\right].\label{gii}\end{equation}

\subsection{Effect of lattice spacing}

\label{MODEDIFF}

Labeling $\alpha$ or $\beta$ as $z$, it was found\cite{paperA}
that for a uniform gas of density $\rho$ with volume $\Delta V$
per lattice point in a $\mc{D}$-dimensional lattice one has \begin{equation}
\frac{\var{\,|\delta^{\textrm{kinetic}}z_{\bo{n}}(t)|\,}}{\var{\,|\delta^{\textrm{direct}}z_{\bo{n}}(t)|\,}}=\mc{R}\approx\order\left[\frac{\hbar^{2}t^{2}\pi^{4}}{60m^{2}(\Delta V)^{4/\mc{D}}}\right]\label{varikd}\end{equation}
 at short times (such that the first term in \eqref{vpp} dominates) when using the standard positive P equations. Here
$\delta^{\textrm{direct}}z_{\bo{n}}$ are the fluctuations in $z_{\bo{n}}$
due directly to the noise in the equations, and present with zero
coupling $\omega_{\bo{nm}}=0$, whereas $\delta^{\textrm{kinetic}}z_{\bo{n}}$
are the remaining fluctuations induced by coupling between modes.
The short time assumption was $\var{\log|z|}\ll1$, which is approximately
$\kappa t\cosh2g''\ll2$, from the formal single mode solutions (31) 
A calculation that takes into account nonzero $g''$ is found to
give the same expression \eqref{varikd}. 

The single mode analysis is expected to be accurate only when this
ratio $\mc{R}\ll1$. For this reason, the simulation time improvements
due to the local diffusion gauge \eqref{gii} are only to be expected
while $\mc{R}\lesssim1$, or perhaps even $\mc{R}\ll1$. Let us see
what this translates to for the case for a uniform gas. 

For a lattice derived from a continuum model,
it will be convenient to write the lattice interaction strength $\kappa=g/\hbar\Delta V$
in terms of the healing length \begin{equation}
\xi^{\textrm{heal}}=\frac{\hbar}{\sqrt{2m\rho g}}.\label{heallength}\end{equation}
 This is the minimum length scale over which a local density inhomogeneity
in a Bose condensate wavefunction can be in balance with the quantum
pressure due to kinetic effects\cite{healinglength}. 

In terms of $\xi^{\textrm{heal}}$, the expected simulation time with
diffusion gauges \eqref{gii} is $t_{\textrm{sim}}\approx\order(10)m\sqrt{\ba{n}}(\xi^{\textrm{heal}})^{2}/\hbar$
from \eqref{giitsim}, where $\ba{n}$ is the mean mode occupation.
The expression \eqref{varikd} can then be evaluated at $t_{\textrm{sim}}$.
Imposing $\mc{R}\ll1$ for times $t\leq t_{\textrm{sim}}$ leads to
the condition $\ba{\Delta\textrm{x}}\gtrsim\xi^{\textrm{heal}}\ba{n}^{1/4}\order(1)$
where the quantity $\ba{\Delta x}=(\Delta V)^{1/\mc{D}}$ is the geometric
mean of the lattice spacing. Local gauges only lead to significant
simulation time improvements in the single mode when $\ba{n}\gg1$. 

We conclude that \textbf{local diffusion gauges will lead to simulation
time improvements when} \begin{equation}
\ba{\Delta\textrm{x}}\gtrsim\order(\xi^{\textrm{heal}}).\label{dxheal}\end{equation}
 The calculations of Section~\ref{MMODE} will be seen to be consistent
with this. This is a strong limitation on local diffusion gauges,
since interesting dynamical effects can readily occur over shorter
length scales than this. It suggests that nonlocal diffusion gauges
may be more useful for general spatially-varying problems, but these
are outside the scope of the present article.

\section{Local drift gauges}

\label{DRIFT} A complementary approach to the diffusion gauge is
to remove the source of the drift instability in Eqn.~(\ref{eq:instability})
by using drift gauges that directly alter the unstable drift equations.
Since the nonlinearity takes the local form $d\alpha_{\bo{n}}=-i\kappa n_{\bo{n}}\alpha_{\bo{n}}\, dt+\dots$
etc., this can be done using drift gauges $\mc{G}_{k}$ dependent
only on local parameters. These methods are certainly capable of
eliminating movable singularities, which are a common cause of 
systematic boundary term errors. However, this is accompanied by
a corresponding growth in sampling error, due to the introduction
of a weight term in the stochastic equations.

\subsection{Single mode}

For a single mode with imaginary diffusion gauge as in Section~\ref{DIFF},
the Ito equations are \begin{eqnarray}\label{1modegaequations}
\fl d\alpha  = & \alpha[-i\kappa n-\gamma/2-i\sqrt{i\kappa}(\mc{G}_{1}\cosh g''+i\mc{G}_{2}\sinh g'')\,]\, dt
+i\sqrt{i\kappa}\alpha\,[dW_{1}\cosh g''+idW_{2}\sinh g''] \nonumber\\
\fl d\beta  = & \beta[i\kappa n-\gamma/2-\sqrt{i\kappa}(-i\mc{G}_{1}\sinh g''+\mc{G}_{2}\cosh g'')\,]\, dt +\sqrt{i\kappa}\beta\,[-idW_{1}\sinh g''+dW_{2}\cosh g'']\nonumber\\
\fl d\Omega  = & \Omega\sum_{k=1}^{2}\mc{G}_{k}dW_{k}.\end{eqnarray}

\subsubsection{Drift gauge}

When $\mc{G}_{j}=0$, the rapid increase in phase variable variance
is due to the process outlined in point 3 of Section~\ref{USEFUL}.
A drift gauge that interrupts this process by removing the offending
drift terms $d\log|\alpha|=\kappa n''\, dt$ and $d\log|\beta|=-\kappa n''\, dt$
is \begin{equation}
\mc{G}_{1}=i\mc{G}_{2}=-\sqrt{i\kappa}e^{-g''}n''.\label{driftgauge1}\end{equation}

\subsubsection{Logarithmic variances}

There is a price paid, of course, and this is fluctuations in the
global weight $\Omega$, such that $\var{\log|\Omega|}$ now scales
$\propto e^{-2g''}$. With a fluctuating weight, quantum phase correlations
such as $G^{(1)}$ are now given by \begin{equation}
G^{(1)}(0,t)=\beta_{0}\langle\Omega\alpha(t)\rangle_{\textrm{s}}=\alpha_{0}^{*}\langle\Omega\beta(t)\rangle_{\textrm{s}}^{*},\label{G1expr}\end{equation}
 using (\ref{moments}). Note that since from \eqref{itodOmega} by
the properties of Ito calculus $d\langle\Omega\rangle_{\textrm{s}}=0$.
Thus the denominator $\langle\Omega+\Omega^{*}\rangle_{\textrm{s}}$
appearing in (\ref{moments}) can be ignored if the initial distribution
is normalized so that $\langle\Omega\rangle_{\textrm{s}}=1$. From
(\ref{G1expr}), the relevant random variables in phase-dependent
calculations are $(\Omega\alpha)$ and $(\Omega\beta)$, so that the
log-phase quantity analogous to $\mc{V}$ in the calculations of Section~\ref{DIFF}
is now \begin{equation}
\mc{V}_{\Omega}=\frac{1}{2}(\var{\log|\Omega(t)\alpha(t)|}+\var{\log|\Omega(t)\beta(t)|})\end{equation}
 (compare to (\ref{mcVdef})). Proceeding as before by formally solving
\eqref{1modegaequations} we find that for off-diagonal coherent state
initial conditions,\begin{equation}
\fl\mc{V}_{\Omega}  =  \frac{\kappa}{2}\left\{t\cosh(2g'')-e^{-2g''}|n_{0}|^{2}\left(\frac{e^{-qt}-1}{q}\right)
-e^{-2g''}[(n'_{0})^{2}-(n''_{0})^{2}]\left(\frac{1-e^{-2\gamma t}}{2\gamma}\right)\right\}\,\:.\label{vga}\end{equation}

Just as in the pure diffusion gauge case, there is a trade-off between
fluctuations due to direct phase variable noise $\propto\cosh(2g'')$
(first term), and fluctuations in the global weight $\Omega$, which
are dependent on $e^{-2g''}$ (following terms). The optimum $g''=g''_{\textrm{opt}}$
can be calculated by solving $\partial\mc{V}_{\Omega}/\partial g''=0$.

\subsubsection{Approximate optimum diffusion gauge}

Since the aim is a gauge adaptable to changing mode occupation like
\eqref{gii}, it is desirable to obtain an approximate expression
for $g''_{\textrm{opt}}$ that can be rapidly evaluated at each time
step in a calculation. To do this, we first consider some special
cases: 

At \textbf{short times} such that $|q|t_{\textrm{opt}}\ll1$ and $2\kappa t_{\textrm{opt}}e^{-2g''}\ll1$
the optimum is given by the roots of the cubic in $V_{g}=e^{-2g''_{\textrm{opt}}}$:
\label{vacubic}\begin{equation}
4\kappa t_{\text{opt}}|n_{0}|^{2}V_{g}^{3}+a_{3}(n_{0},\gamma t_{\textrm{opt}})V_{g}^{2}-1=0\end{equation}
 where \begin{equation}
a_{3}(n_{0},v)  =  1+4(n''_{0})^{2}\left(\frac{1-e^{-2v}}{2v}\right)
 -2|n_{0}|^{2}\left(\frac{1-2v+2v^{2}-e^{-2v}}{v}\right).\quad\label{a3def}\end{equation}
 For zero damping $a_{3}(n_{0},0)=1+4(n''_{0})^{2}$. 

In the case of simulations with sizable occupation of modes, and times
up to $\order(t_{\textrm{coh}})$ the short time conditions are satisfied
and this cubic applies. When the $V_{g}^{2}$ term is negligible we
have \begin{equation}
g''_{\text{opt}}\approx\frac{1}{3}\log\left(|n_{0}|\sqrt{4\kappa t_{\text{opt}}}\right).\label{approxgiiopt1}\end{equation}
 This occurs when $n_{0}$ is large enough and mostly real, and $t$
is big enough: i.e. when $a_{3}(n_{0},\gamma t)\ll(4\kappa t_{\text{opt}}|n_{0}|^{2})^{2/3}$.
For example, when undamped, $\kappa t_{\text{opt}}$ must be at least
$\gg1/4|n_{0}|^{2}$ with classical initial conditions ($n''_{0}=0$).
The opposite case when $n_{0}$ is either too small, too imaginary,
or the time is too short has the $V_{g}^{3}$ term negligible and
leads to \begin{equation}
g''_{\text{opt}}\approx\frac{1}{4}\log a_{3}(n_{0},\gamma t_{\textrm{opt}}).\label{approxgiiopt2}\end{equation}

For strong damping $q>0$, we again have linear growth $\mc{V}_{\Omega}=\kappa t\cosh(2g'')/2-b_{2}$,
where the constant is now \begin{equation}
b_{2}=\frac{\epsilon}{4}\{[(n'_{0})^{2}-(n''_{0})^{2}]e^{-2g''}-|n_{0}|^{2}/(e^{2g''}-\epsilon)].\end{equation}

\begin{figure}
\center{\includegraphics[width=250pt]{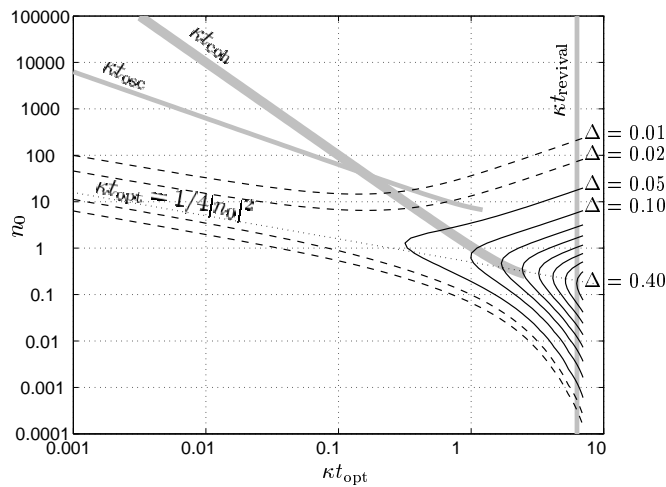}}
\caption{\label{FIGgadisc} 
Discrepancy $\Delta=g''_{\text{opt}}-g''_{\text{approx}}$ between
$g''_{\textrm{opt}}$ (the exact optimization of $g''$ from (\ref{vga})
and the approximate expression (\ref{noptga}) for an undamped mode
with diagonal coherent initial occupation $n_{0}=n'_{0}$. Discrepancy
values $\Delta$ are shown as contours. Dotted line approximates region
of greatest discrepancy $\kappa t_{\textrm{opt}}\approx1/4|n'_{0}|^{2}$. }
\end{figure}

An approximation for the diffusion gauge that is found to work well
in practice (see Section~\ref{1MODE}) is \begin{equation}
g''_{\text{approx}}=\frac{1}{6}\log\left\{ 4|n_{0}|^{2}\kappa t_{\text{opt}}+a_{3}(n_{0},\gamma t_{\textrm{opt}})^{3/2}\right\} ,\label{noptga}\end{equation}
 which reduces to (\ref{approxgiiopt1}) and (\ref{approxgiiopt2})
in their limits of applicability. The discrepancy $\Delta$ between
(\ref{noptga}) and the exact optimization obtained by solving $\partial\mc{V}_{\Omega}/\partial g''=0$
is shown for real $n_{0}$ in Figure~\ref{FIGgadisc}.

\subsubsection{Useful simulation time}

Consider the undamped high occupation regime with coherent state initial conditions $n_{0}=n'_{0}\gg1$. Using \eqref{approxgiiopt1},
at target times $t_{\rm opt}\gg1/4\kappa n_0^2$, one has
$e^{-2g''_{\textrm{approx}}}\approx1/[4n_{0}^{2}\kappa t_{\rm opt})^{1/3}\ll1$.
Then the terms in \eqref{vga} of highest order in $n_0$ give
$\mc{V}_{\Omega}(t=t_{\rm opt})\approx \frac{3}{2}[(\kappa t_{\rm opt})^2n_0/4]^{2/3}$.
(Target times of interest will almost always satisfy the prior condition.)
We can use the condition $\mc{V}_{\Omega}\lesssim10$ from \eqref{varflimit} 
 to estimate the useful simulation time in this regime as \begin{equation}
t_{\textrm{sim}}\approx\order(10)t_{\textrm{coh}}.\label{tsimga}\end{equation}
 This is again a large improvement compared to the positive P results  in Table~\ref{TABLEtsim}.

\subsection{Extension to many modes}

\label{GAUGEMMODE}

We now wish to consider how the drift gauge approach can be used to
treat a multi-mode situation.

\subsubsection{Adaptive gauge}

Proceeding to as in Section~\ref{ADAPT}, the suggested gauges with
the present approach are: \begin{equation}
g''_{\bo{n}}=\frac{1}{6}\log\left\{ 4|n_{\bo{n}}(t)|^{2}\kappa t_{\textrm{rem}}+a_{3}(n_{\bo{n}}(t),\gamma t_{\textrm{rem}})^{3/2}\right\} ,\label{giiga}\end{equation}
 appearing in noise matrices of the form \eqref{Bdef}, and \begin{equation}
\mc{G}_{j}=i\mc{G}_{j+M}=-\sqrt{i\kappa}\text{Im}[n_{\bo{n}_{j}}(t)]\exp\left(-g''_{\bo{n}_{j}}(t)\right).\label{driftgauge}\end{equation}
 for $j=1,\dots,M$.

\subsubsection{Drift gauges and weight spread}

\label{MODEDRIF}

Drift gauged simulations using \eqref{driftgauge} encounter a scaling
problem in many-mode systems because the single weight variable $\Omega$
accumulates fluctuations from all modes (see \eqref{itodOmega}).
There are precisely two independent noises and two drift gauges for
each mode. Consider for example, a uniform gas of density $\rho$,
and volume $V$ on $M$ modes, so that the mean mode occupation is
$\ba{n}=\rho\Delta V$ Writing $\mc{G}_{k}=\mc{G}'_{k}+i\mc{G}''_{k}$
and $\log\Omega=\theta'+i\theta''$, from \eqref{itodOmega} one can
show that \begin{equation}
\frac{d}{dt}\var{\theta'}=\sum_{k}\langle(\mc{G}'_{k})^{2}+\text{covar}[\theta',(\mc{G}''_{k})^{2}]-\text{covar}[\theta',(\mc{G}'_{k})^{2}]\rangle_{\textrm{s}}.\label{logweightspread}\end{equation}
 For the uniform gas the contribution from each mode is identical
on average, and from \eqref{driftgauge}, $\mc{G}'_{k}$ and $\mc{G}''_{k}$
are of the form $\pm\sqrt{\kappa/2}\,e^{-g''}\im{n}$. Hence, approximately,
\begin{equation}
\frac{d}{dt}\var{\log|\Omega|}\propto M\kappa\langle e^{-2g''}\im{n}^{2}\rangle_{\textrm{s}}\,\,.\end{equation}

From the formal solution \eqref{nformalpp} (which also applies with
drift gauges \eqref{driftgauge}), one has $\var{\im{n}}=\langle(|n|\sin\im{\log n})^{2}\rangle_{\textrm{s}}\approx\ba{n}^{2}\langle\sin^{2}\im{\log n(t)}\rangle_{\textrm{s}}$,
which (at short times $\kappa te^{-2g''}\ll1$) is $\approx\ba{n}^{2}\kappa te^{-2g''}$.
Thus at these short times \begin{equation}
\frac{d}{dt}\var{\log|\Omega|}\approx\,\propto M\kappa^{2}\ba{n}^{2}te^{-4g''},\end{equation}
 and at high lattice occupations $\ba{n}\gg1$ when using \eqref{giiga}
$e^{-2g''}\approx1/(4\ba{n}\kappa t)^{1/3}$, so, substituting: \begin{equation}
\var{\log|\Omega|}\approx\,\propto M(g\rho t)^{4/3}.\end{equation}
 Imposing the log variance limit \eqref{varflimit} on $|\Omega|$
(because the factor $|\Omega|$ appears in all observable estimates
\eqref{moments}), one obtains the estimate that \begin{equation}
t_{\textrm{sim}}\approx\,\propto\frac{1}{g\rho M^{3/4}}.\end{equation}

For this reason we expect that simulations using the local diffusion
and local drift gauges \eqref{driftgauge} and \eqref{giiga} will
only give significant simulation time improvements when the number
of highly occupied modes is relatively small. Indeed, the two mode
cases of Section~\ref{2MODE} show strong simulation time improvement
with this gauge, while we have found that the many-mode uniform gases of Section~\ref{MMODE}
do not simulate well with the gauge form \eqref{driftgauge}. 

Note, however, that since single-mode $t_{\textrm{sim}}$ decreases
rapidly with occupation $n$, it is the most highly occupied modes
that limit the simulation time. So, even a very large $M$ system
may still experience improvements in simulation time with the present
method if there are only a few modes with the highest occupations. 

As in the pure diffusion gauge case, this does not preclude that better
scaling may be obtainable with appropriately tailored \textit{nonlocal}
drift and diffusion gauges. In particular, the local drift gauge employed
here does not take into account the fact that neighbouring lattice
points with spacings of less than a healing length become strongly
correlated due to the kinetic energy terms. These cause particle exchange,
and hence effectively average out local fluctuations in $n''$.

\section{Single mode: numerical results}

\label{1MODE} Simulations of an undamped single-mode anharmonic oscillator
(see Section~\ref{SINGMODE}) were performed for a wide range of
initial coherent states $n_{0}=n'_{0}$ from $10^{-5}$ to $10^{10}$.
Results for the standard positive P method were reported in \cite{paperA}.
Here the gauged methods described in Sections~\ref{DIFF} and~\ref{DRIFT}
are tested.

\subsection{Procedure}

In what follows, the term \textit{useful precision} for an observable
$\op{O}$ has been taken to indicate the situation where the estimate
$\ba{O}=\langle f\rangle_{\textrm{s}}$ of $\langle\op{O}\rangle$
using $\mc{S}=10^{6}$ trajectories has a relative precision of at
least $0.1$ at the one sigma confidence level. This is assuming the
CLT holds so that \begin{equation}
\Delta\ba{O}\approx\sqrt{\frac{\var{f}}{\mc{S}}}\end{equation}
 is used to assess uncertainty in $\ba{O}$. For the model here, we
consider useful precision in the magnitude of phase-dependent correlations
$|G^{(1)}(0,t)|$, which is the low-order observable most sensitive
to the numerical instabilities in the equations. 

Uncertainties in the calculated useful times $t_{\textrm{sim}}$ arise
because the $\Delta|G^{(1)}|$ were themselves estimated from finite
ensembles of $\mc{S}=10^{4}$ trajectories. The range of $t_{\textrm{sim}}$
indicated in Figure~\ref{FIGtsim} was obtained from 10 independent
runs with identical parameters. 

Taking the analytic scalings (\ref{giitsim}) \eqref{tsimga} at high
$n_{0}$, and from Ref.~\cite{paperA} at low $n_{0}$ into account,
parameters in an approximate curve \begin{equation}
t_{\text{est}}=\frac{1}{\kappa}\left\{ \left[{c_{1}}(n'_{0}){}^{-c_{0}}\right]^{-{c_{2}}}+\left[2\log\left(\frac{e^{c_{3}}}{n'_{0}{}^{c_{4}}}+1\right)\right]^{-{c_{2}}}\right\} ^{-1/{c_{2}}}\label{timefit}\end{equation}
 have been fitted to the empirical data, as was done for the standard
positive P method in \cite{paperA}. Best values of $c_{j}$ are given
in Table~\ref{TABLEtsimfit}. The expression (\ref{timefit}) reduces
to ${c_{1}}n'_{0}{}^{-c_{0}}/\kappa$ and $({c_{3}}-c_{4}\log n'_{0})/(\kappa/2)$,
when $n'_{0}\gg1$ and $n'_{0}\ll1$, respectively. $c_{0}$ determines
the high $n'_{0}$ scaling (here, assumed from analysis of $\mc{V}$,
or $\mc{V}_{\Omega}$), $c_{1}$ characterizes the pre-factor for
high $n'_{0}$, $c_{3}$ is proportional to a constant residual $t_{\textrm{sim}}$
at near vacuum, $c_{4}$ characterizes the curvature at small $n'_{0}$,
while $c_{2}$ is related to the stiffness of the transition between the two regimes. Uncertainty
$\Delta c_{j}$ in parameters $c_{j}$ was worked out by requiring
$\sum_{n_{0}}\{[t_{\text{est}}(c_{j}\pm\Delta c_{j},n_{0})-t_{\text{sim}}(n_{0})]/\Delta t_{\text{sim}}\}^{2}=\sum_{n_{0}}\{1+([t_{\text{est}}(c_{j},n_{0})-t_{\text{sim}}(n_{0})]/\Delta t_{\text{sim}})^{2}\}$. 

For calculations involving a diffusion gauge dependent on target time
$t_{\textrm{opt}}$, a wide variety of target times were tried to
investigate the dependence between $t_{\textrm{sim}}$ and $t_{\textrm{opt}}$,
and ascertain what are the longest simulation times achievable.

\subsection{Simulation times}

\begin{figure}
\center{\includegraphics[width=300pt]{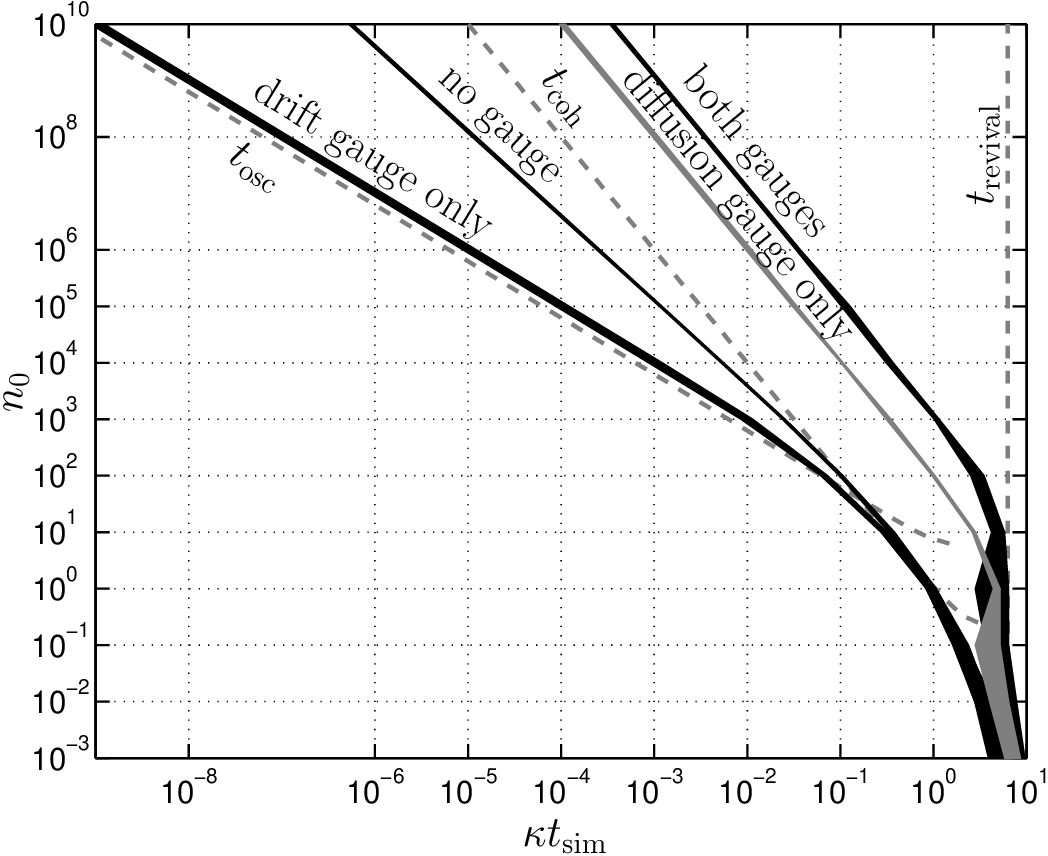}}\vspace*{-10pt}
\caption{\label{FIGtsim} Maximum useful simulation time $t_{\textrm{sim}}$,
of the one-mode undamped anharmonic oscillator with various gauge
choices. Initial coherent state mean mode occupations $n_{0}=n'_{0}$.
Width of plotted lines shows range, when using 10 runs of $\mc{S}=10^{4}$
trajectories each. The drift gauge is (\ref{driftgauge}), while the
diffusion gauge is (\ref{gii}) when on its own, or (\ref{giiga})
when combined with the drift gauge. }
\end{figure}

\begin{figure}
\center{\includegraphics[width=\columnwidth]{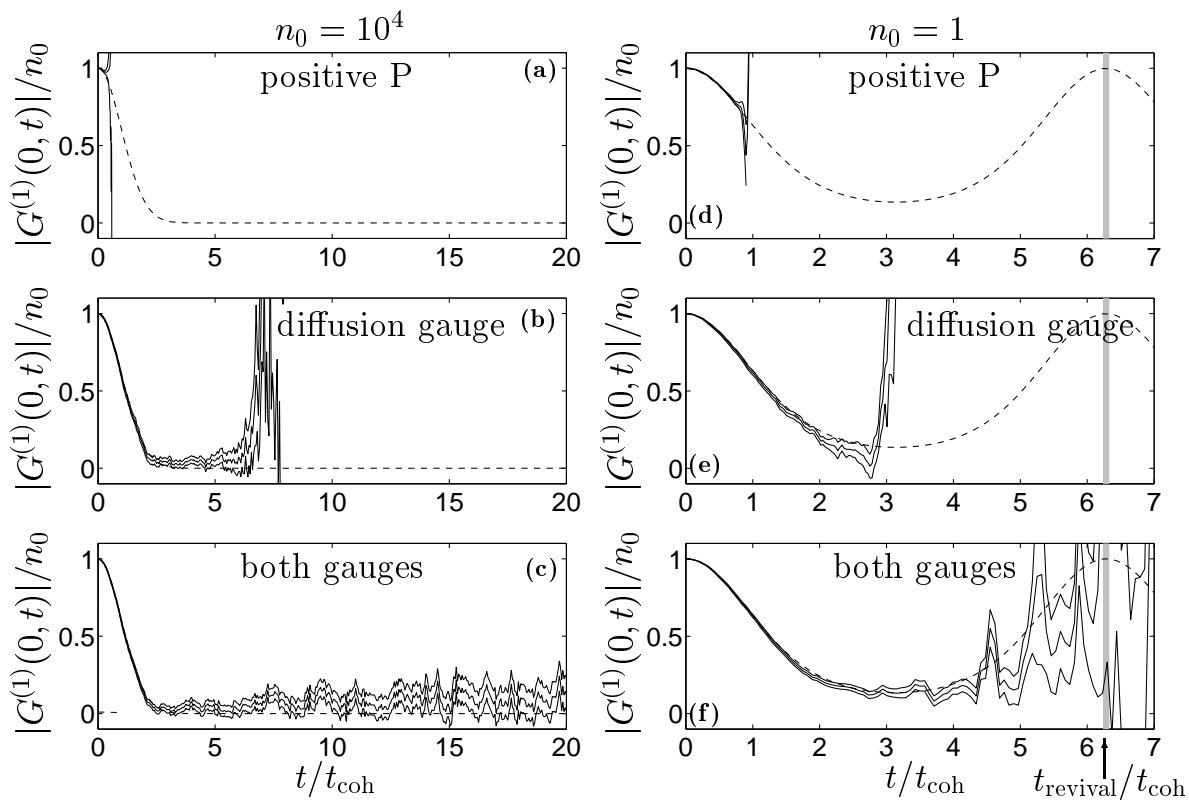}}\vspace*{-10pt}
\caption{\label{FIGG} 
Modulus of the phase correlation function $G^{(1)}(0,t)$. Comparison
of calculations with different gauges. Subplots \textbf{(b)} and \textbf{(e)}:
use diffusion gauge \eqref{plimakgauge} of Plimak\etal\cite{noiseoptimisation}
with $t_{\textrm{opt}}=3t_{\textrm{coh}}$, while subplots \textbf{(c)}
and \textbf{(f)}use the combined drift and adaptive diffusion gauges
(\ref{driftgauge}) and (\ref{giiga}) with the choice $t_{\text{opt}}=20\, t_{\text{coh}}$.
The initial conditions: coherent state with $\langle\op{n}\rangle=n_{0}$.
Triple solid lines indicate $G^{(1)}$ estimate with error bars, dashed
lines are exact values. $\mc{S}=10^{4}$ trajectories in all cases.}
\end{figure}

Figure~\ref{FIGtsim} compares $t_{\textrm{sim}}$, as defined via ``useful precision'' in $|G^{(1)}(0,t)|$,   for several gauge
choices. Note that a logarithmic scale is used. Results at high occupation
are tabulated in Table~\ref{TABLEtsim}, which includes data for
a larger set of gauge choices. Figure~\ref{FIGG} gives examples
of calculated values $|G^{(1)}(0,t)|$ along with error estimates.
Table~\ref{TABLEtsimfit} gives empirical fitting parameters to the
expression (\ref{timefit}). 

\newcommand{\mm}[2]{{}^{#1}_{#2}}%
\begin{table}
\caption{\label{TABLEtsim} Some properties of simulation times $t_{\textrm{sim}}$
for various gauges when applied to the undamped one-mode anharmonic
oscillator. Initial coherent state mean occupation $n_{0}=n'_{0}$. }
\lineup
\begin{indented}\item[]\begin{tabular}{@{}ccc@{(\ }l@{$\ \pm\ $}l@{\ )\ }lll}\br
Drift gauge 		& Diffusion gauge	&\multicolumn{4}{c}{Useful simulation time $t_{\rm sim}$}
												& \multicolumn{2}{c}{Maximum $n_0$}\\
$\mc{G}_k$		& $g''$			&\multicolumn{4}{c}{maximized over $t_{\rm opt}$,}
												& \multicolumn{2}{c}{for which}\\
			&			&\multicolumn{4}{c}{when $n_0=n'_0\gg1$}	& \multicolumn{2}{c}{$\chi t_{\rm sim}\ge1$}\\
\mr
\eqref{CCDgauge}	& 0			&& \01.06 & 0.16	& $t_{\text{osc}}$		
												&\0\00.014 &$\mm{+\ \ \00.016}{-\ \ \00.008}$\\
\eqref{driftgauge}	& 0			&& \01.7 & 0.4		& $t_{\text{osc}}$	
												&\0\00.08 &$\mm{+\ \ \00.07}{-\ \ \00.05}$\\
\bf 0			& \bf 0			&&\bf \02.5 & \bf 0.2	&$t_{\text{coh}}/n'_0{}^{1/6}$	
												&\0\00.11 &$\pm\ \00.06$\\
0			&\eqref{plimakgauge} or (36)			
						&& \08.2 & 0.4		& $t_{\text{coh}}$ 	&\012 &$\pm\ \03$\\
0			&\eqref{gii}		&& 10.4 & 0.7		& $t_{\text{coh}}$ 	&\019 &$\pm\ \04$\\
\eqref{driftgauge} 	&\eqref{noptga}		&& 30 & 3		& $t_{\text{coh}}$ 	&150 &$\pm\ 40$\\
\textbf{\eqref{driftgauge}}
			&\textbf{\eqref{giiga}} &&$\!$\bf35 & \bf 4 	& $t_{\text{coh}}$ 	&240 &$\pm\ 70$\\
\br
\end{tabular}\end{indented}
\end{table}

\begin{table}
\caption{ \label{TABLEtsimfit}
Empirical fitting parameters for maximum useful
simulation time $t_{\textrm{sim}}$ with several different gauge choices
when applied to the undamped one-mode anharmonic oscillator. The fit is to expression (\ref{timefit}).
}
\lineup
\begin{indented}\item[]\begin{tabular}{ll@{$\ \pm\ $}ll@{$\ \pm\ $}ll@{$\ \pm\ $}ll@{$\ \pm\ $}l}\br
	& \multicolumn{2}{l}{Positive P}&\multicolumn{2}{l}{Drift gauge}	&\multicolumn{2}{l}{Diffusion gauge} 
										  &\multicolumn{2}{l}{Both gauges} \\
\mr
$G_k$ 	& \multicolumn{2}{c}{0} 	& \multicolumn{2}{c}{(\ref{driftgauge})}&\multicolumn{2}{c}{0}
										  & \multicolumn{2}{c}{(\ref{driftgauge})}\\
$g''$	& \multicolumn{2}{c}{0}		&\multicolumn{2}{c}{0} 			&\multicolumn{2}{c}{\eqref{gii}}
										  & \multicolumn{2}{c}{(\ref{giiga})} \\
${c_0}$	& \multicolumn{2}{c}{$2/3$}	&\multicolumn{2}{c}{$1$}		&\multicolumn{2}{c}{$1/2$}
										  &\multicolumn{2}{c}{$1/2$}	\\
\br
${c_1}$	& $\m2.5$	& $0.2$		& $\,11$		&$3$ 		&$10.4$		&$0.7$	
										& $35$		& $4$		\\
${c_2}$	& $\m3.2$	&$\mm{\infty}{1.2}$
					& $\m1.4$		& $\mm{\infty}{0.4}$
									 &$\02.7$		&$\mm{\infty}{1.0}$
										& $\03.6$		& $\mm{\infty}{2.3}$	\\
${c_3}$	& $-0.5$	& $0.3$		& $-0.5$	& $0.3$ 	&$\02.4$		&$0.6$	
										& $\02.8$		& $0.9$		\\
${c_4}$	& $\m0.45$	& $0.07$	& $\m0.49$	& $0.08$	&$\00.23$		&$0.13$	
										& $\00.23$	& $0.13$	\\
\br
\end{tabular}\end{indented}
\end{table}

We see that: 

\begin{itemize}
\item Combining drift and diffusion gauges gives the longest useful simulation
times. Such simulations give good precision well beyond the point
at which all coherence has decayed away for highly-occupied modes
- potentially up to about 35 collapse times $t_{\rm coh}$ in the cases treated
here. 
\item Diffusion-gauge-only simulations also give quite good statistical
behaviour (although useful simulation times are about 4 times shorter
at high occupation than with both gauges). 
\item Despite the efficient behaviour of combined gauge simulations, using
only a drift-gauge gives even worse statistical error than no gauge
at all. Such simulations are restricted in time to about one phase
oscillation. This indicates that the drift gauge choice made here
is not an optimum choice when used in isolation.
\item At low occupation, i.e. of the order of one boson or less, combined
gauge methods still give the best results, but the advantage is marginal. 
\item The simulation times with nonzero diffusion gauges (whether accompanied
by drift gauge \eqref{driftgauge} or not) not only have better scaling
with $n_{0}$ when $n_{0}$ is large, but this power-law decay of
simulation time starts much later, as seen in Figure~\ref{FIGtsim}
and the right hand column of Table~\ref{TABLEtsim}. 
\end{itemize}

\subsection{Target time dependence}

\label{TARGETT} 
\begin{figure}
\center{\includegraphics[width=400pt]{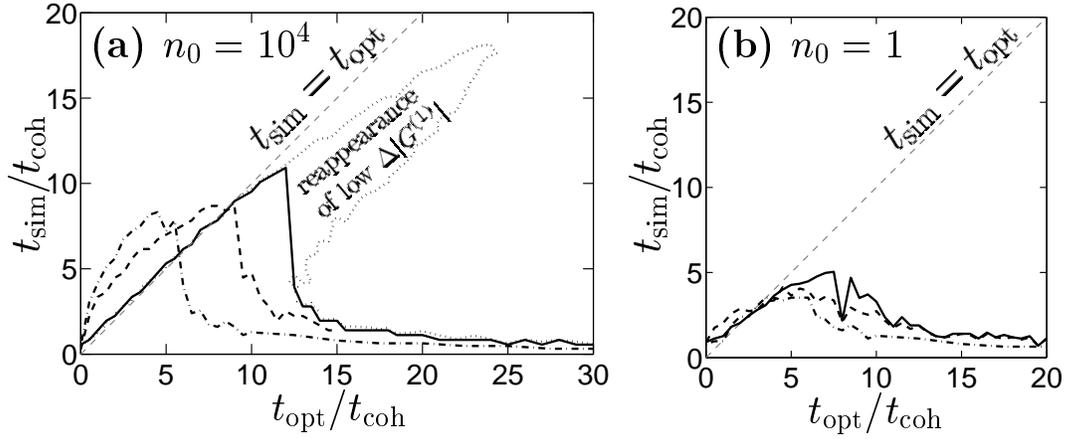}}\vspace*{-10pt}
\caption{\label{FIGtoptnopt} Dependence of actual useful simulation time
$t_{\textrm{sim}}$ on the \textit{a priori} target time $t_{\textrm{opt}}$
for a variety of diffusion gauges in the $\mc{G}_{k}=0$ schemes:
Solid line: full adaptive gauge \eqref{gii}; dashed line: number-adaptive-only
gauge of form (36) but with the modification $n_{0}\rightarrow n(t)$;
dash-dotted line: gauge \eqref{plimakgauge} of Plimak\etal\cite{noiseoptimisation}.
For the gauge (36), the whole region where useful precision
occurs is shown by the \textsc{dotted} line. Undamped anharmonic oscillator
system \eqref{1modeH}, with coherent state initial conditions $n_{0}=n'_{0}$. }
\end{figure}

\begin{figure}
\center{\includegraphics[width=400pt]{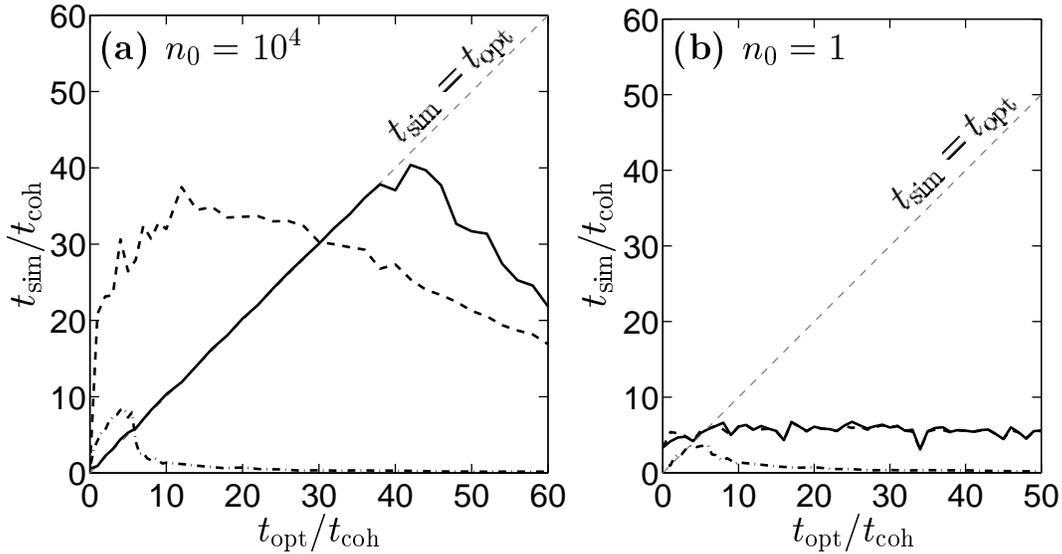}}\vspace*{-10pt}
\caption{\label{FIGtopt} Dependence of $t_{\textrm{sim}}$ on $t_{\textrm{opt}}$,
as in Figure~\ref{FIGtoptnopt}, with drift gauge \eqref{driftgauge}
as well as diffusion gauges. Details as in Figure~\ref{FIGtoptnopt}. }
\end{figure}

Figures~\ref{FIGtoptnopt} and ~\ref{FIGtopt}
show the dependence of simulation times on the target time parameter
$t_{\textrm{opt}}$ for a variety of gauges. Comparison is made between
the full adaptive forms $g''(n_{\bo{n}}(t),t_{\textrm{rem}})$ (solid
lines), number-adaptive-only forms $g''(n_{\bo{n}}(t),t_{\textrm{opt}})$
that do not use explicit time-dependence $t_{\textrm{rem}}$ (dashed
lines), and the Plimak\etal\cite{noiseoptimisation} gauge \eqref{plimakgauge}
(dash-dotted lines). 

Some comments: 

\begin{itemize}
\item The diffusion gauge forms that optimize for the {}``remaining time
to target'' $t_{\textrm{rem}}$ give the longest simulation times,
and these times are well controlled. Statistical error can be reliably
expected to remain small up to the explicit target time $t_{\text{opt}}$,
provided that this is within the useful simulation range given in
Table~\ref{TABLEtsim}. 
\item Diffusion gauges that instead use a constant $t_{\textrm{opt}}$
parameter give: 

\begin{enumerate}
\item Somewhat shorter simulation times. 
\item A complicated relationship between target and useful simulation times.
Broadly speaking $t_{\textrm{opt}}\ll t_{\textrm{sim}}$ for the optimum
cases. 
\end{enumerate}
These forms of diffusion gauges require tedious parameter searching
to find the best $t_{\textrm{opt}}$ choice for given initial conditions.
The probable reason why $t_{\textrm{sim}}\neq t_{\textrm{opt}}$ here
is that $g''$ has been optimized to minimize the variance of \textit{logarithmic}
variables. This is not the same as the variance of the non-logarithmic
variables $\alpha$, or $\beta$ that actually appear in the observable
calculations. Hence, different $g''$ forms may extend simulation
time by a further amount. 

\item When drift gauges \eqref{driftgauge} are used, an adaptive diffusion
gauge $g''(n(t))$ rather than a constant $g''(n_{0})$ can give much
longer simulation times. (Compare to the Plimak\,\etal gauge in Figure~\ref{FIGtopt}\textbf{(a)}
which effectively scans through a range of constant $g''$ values). 
\item At low $n'_{0}$, $t_{\textrm{sim}}$ is only weakly dependent on
local diffusion gauge choice. 
\item When there is no drift gauge, for $n'_{0}\gg1$, the time-adaptive
gauge forms $g''(t_{\textrm{rem}})$ lead to a peculiar effect if
the optimization time $t_{\textrm{opt}}$ is chosen larger than the
usual maximum $t_{\textrm{sim}}$ given in Table~\ref{TABLEtsim}.
The statistical error in the $G^{(1)}$ estimate first rises rapidly,
then falls again, and finally grows definitively. The parameter region
in which this occurs is shown in Figure~\ref{FIGtoptnopt}\textbf{(a)}.
In effect one has two time intervals when the simulation gives useful
results: at short times, and later in a time interval around $t\approx\order(10t_{\textrm{coh}})$. 
\end{itemize}
More detail on these numerical investigations can be found in Ref.~\cite{inprep},
Chapter 7.

\subsection{Comparison to recent related work}

\label{OTHERGAUGE}

Improvements to the basic positive P simulation method for specific
cases of interacting Bose gas systems have been tried with some success
in several recent publications\cite{carusottodynamix,noiseoptimisation,ccp2k,stochasticgauges}.
Here we compare these with the stochastic gauge formalism, and make
some comparison to the results and analysis in the present chapter.

\subsubsection{The work \cite{carusottodynamix} of Carusotto, Castin, and Dalibard}

\label{CH7ComparisonCCD} An isolated (i.e. particle-conserving) system
of exactly $\ba{N}$ interacting bosons was considered (on a 1D lattice).
The {}``coherent state simple scheme'' for stochastic wavefunctions
described in Section III B 2 therein can be identified as using drift
gauges of the form \begin{eqnarray}\eqalign{\label{CCDgauge}
\mc{G}_{1}  = &i\sqrt{i\kappa}\left(n-|\alpha|^{2}\right) \\
\mc{G}_{2}  = &\hspace*{1ex}\sqrt{i\kappa}\left(n-|\beta|^{2}\right)}\end{eqnarray}
 when re-written our notation for the single-mode system \eqref{1modeH}.
This gauge causes a full decoupling of the complementary $\alpha$
and $\beta$ equations by making the replacements $n\rightarrow|\alpha|^{2}$
or $n\rightarrow|\beta|^{2}$ in the nonlinear terms. Like \eqref{driftgauge}
it is also successful in removing movable singularities, since the
nonlinear terms in the radial equations for $d|\alpha|$ and $d|\beta|$
are removed. 

The stochastic wavefunction in the form presented in \cite{carusottodynamix}
is only applicable to closed systems with a definite and explicit
number of particles, so e.g. evaporative cooling or coherent out-coupling
from a system cannot be treated.

\subsubsection{The work \cite{noiseoptimisation} of Plimak, Olsen, and Collett}

\label{CH7ComparisonPlimak}

A single-mode undamped, gainless system \eqref{1modeH} at high Bose
occupation with coherent state initial conditions was considered.
The {}``noise optimization'' scheme applied therein to greatly improve
simulation times can be identified as an imaginary diffusion gauge
of the form (rewritten in the present notation) \begin{equation}
g''=g''_{A}=\frac{1}{2}\cosh^{-1}\left[n_{0}\kappa t_{\textrm{opt}}\right]\label{plimakgauge}\end{equation}
 defined at high occupation or long times (i.e. while $n_{0}\kappa t_{\textrm{opt}}\geq1$).
This is dependent on a target time $t_{\textrm{opt}}$ (which was
taken to be $t_{\textrm{opt}}=3t_{\text{coh}}$ in the calculations
of Ref.~\cite{noiseoptimisation}), and the initial Bose occupation
$n_{0}=n'_{0}$. The useful simulation times obtainable with this
method are also shown in Figures~\ref{FIGtsim} and~\ref{FIGG},
and Table~\ref{TABLEtsim}. Their dependence on $t_{\textrm{opt}}$
has been calculated here, and is shown in Figure~\ref{FIGtoptnopt}.

\subsubsection{The work \cite{stochasticgauges} of Drummond and Deuar}

\label{WIGNER} In section 5.3 of the above article, some preliminary
results for the dynamics of a one-mode, undamped, gainless system
(with $n_{0}=9$ particles on average) were shown. The drift gauge
(\ref{driftgauge}), and a constant imaginary diffusion gauge $g''=1.4$
were used.

\section{Convergence issues}
\label{BT}

\subsection{General}

A subtle issue with many phase-space distributions is the possibility
of so-called boundary term errors. These can arise when the tails
of a distribution (say $G(\alpha,\beta,\Omega)$) do not fall off fast
enough as the boundaries of phase space are approached (in the case
here, as $|\alpha|,|\beta|$, or $|\Omega|\rightarrow\infty$). It
is possible for this to lead to a bias in means of random variables
even in the infinite sample limit, if parts of the distribution 
which have a non-negligible effect are never sampled. 

Some numerical indicators have been developed\cite{boundarytermerrors}
that allow one to search for symptoms of these errors using the numerical
data. The most useful of these indicators is sudden appearance of
spiking in observable estimates, where the onset of this spiking tends to come
earlier as more trajectories are added. Such spikes usually occur when a single trajectory samples a long power-law tail. \textbf{Results obtained after first
spiking are suspect}.

In our experience\cite{inprep}, another indicator can be obtained by performing 
two simulations with sample sizes $\mc{S}$ differing by an order of magnitude. 
If a \textbf{statistically significant difference (e.g. $2\sigma$) in observable predictions} occurs, 
the simulation is suspect.

In all cases with which we are familiar, symptoms of potential convergence problems 
have been apparent already in the equations of motion. They have been of two kinds\cite{inprep}\footnote[1]{Of course, other ways of achieving a divergence may occur}: 

\subsection{Divergence symptoms of the first kind: movable singularities}
\label{2NDKIND}
This is a symptom apparent in the drift parts of the equations of motion.

It has been found\cite{boundarytermerrors} that a systematic boundary term error
is often associated with the presence of so-called \textit{movable
singularities}. These are trajectories (usually of measure zero) in
the deterministic parts of the equations that diverge in a finite
time. The presence of a bias in this situation can be understood by
considering that the effect of an infinitesimal number of divergent
trajectories may be nonzero. 
Some cases where this occurs with a positive P simulation have
been investigated by Gilchrist \textit{et al} \cite{boundarytermerrors}.
The bias has been shown to disappear
when the movable singularities are removed using drift gauges\cite{removalofboundaryterms}. See also Ref.~\cite{inprep}, Chapter 6. 

Consider a set of generic stochastic equations \begin{equation}\label{genericeqn}
dv=A_{v}(\bo{v})dt+\sum_{k}B_{vk}(\bo{v})dW_{k}(t),\end{equation}
 in complex variables $\bo{v}=\{ v\}$ free to explore the whole complex
plane. If, in the limit $|v|\to\infty$, deterministic growth of $|v|$ is exponential or slower, 
then the trajectory cannot reach infinity in a finite time by deterministic processes, and moving singularities are ruled out. Exponential growth in $v$ occurs when $\re{A_v/v}$ is a positive constant, so we conclude that 
a condition sufficient to rule out moving singularities is that
\begin{equation}
\lim_{|v|\rightarrow\infty}\re{\frac{A_{v}}{v}}\label{mvsingc}
\end{equation}
 converges for all variables $v$. 

\subsection{Divergence symptoms of the second kind: noise-weight divergence}
This is a symptom which arises from a combination of the noise behaviour and the form of the quantities which are averaged to obtain observable estimates.

A classic and well known example of this convergence problem occurs for the present single mode system (\ref{1modeH})
when using an un-normalized Bargmann coherent state kernel
\begin{equation}
 \op{\Lambda}(\alpha,\beta) = ||\alpha\rangle\langle\beta^*||.
\end{equation}
(Compare to the gauge P kernel (\ref{kernel})\,). The (non) convergence of this model has been considered by Carusotto and Castin\cite{CCconvergence}, and also in Ref.~\cite{inprep}, Section 6.2.2. 

One finds that the Ito equations of motion are
\numparts\begin{eqnarray}
d\alpha &=& i\alpha\sqrt{i\kappa}\,dW_1\\
d\beta &=& \beta\sqrt{i\kappa}\,dW_2.
\end{eqnarray}\endnumparts
To calculate estimates for an observable $\op{O}$ we need to evaluate 
\begin{equation}
\frac{\tr{\op{O}\op{\rho}}}{\tr{\op{\rho}}} = \frac{\left\langle\tr{\op{\Lambda}\op{O}}\right\rangle_{\rm s}}{\left\langle\tr{\op{\Lambda}}\right\rangle_{\rm s}} = \frac{\left\langle f(\alpha,\beta)\,e^{\alpha\beta}\right\rangle_{\rm s}}{\left\langle e^{\alpha\beta}\right\rangle_{\rm s}}
\end{equation}
for some appropriate $f(\alpha,\beta)$ which depends on the details of $\op{O}$.
The equations of motion can be formally solved to give
\begin{equation}
n(t) = \alpha(t)\beta(t) = n(0)\exp\left[-\sqrt{\kappa}\xi^-(t) +i\sqrt{\kappa}\xi^+(t)\right],
\end{equation}
with the variance $t$ Gaussian random variables $\xi^{\pm}$ defined similarly to (\ref{zetadef}):
\begin{equation}\xi^{\pm} = \frac{1}{\sqrt{2}}\int_{s=0}^{\,t}\left[\,dW_1(s)\pm dW_2(s)\,\right].
\end{equation}
For any observable estimate, we must estimate $\langle e^{\alpha\beta}\rangle_{\rm s}$ using our samples of $n(\xi^+,\xi^-)$. The distribution of this can be explicitly evaluated:
\begin{eqnarray}
\langle e^{\alpha\beta}\rangle_{\rm s} &=& \int_{-\infty}^{\infty} P(\xi^+) P(\xi^-)\ e^{\alpha\beta} d\xi_1\,d\xi_2\nonumber\\
&\propto& \int_{-\infty}^{\infty} \exp\left\{ n(0)\,e^{-\sqrt{k}\xi^-}e^{i\sqrt{k}\xi^+} -\frac{(\xi^-)^2}{2t}-\frac{(\xi^+)^2}{2t}\right\}\, d\xi^-\, d\xi^+.
\end{eqnarray}

This distribution is divergent as $\xi^-\to -\infty$ because the $e^{-\sqrt{k}\xi^-}$ factor in the exponent always beats the (relatively) weak gaussian tail, which can only manage a quadratic $-(\xi^-)^2/2t$ drop-off in the exponent.

\subsection{Stochastic model system}

A more general feeling for the effect on convergence of the noise-weight relationship can be gained by considering the following simple equation for a complex variable $v$:
\begin{equation}
dv(t) = c\  v(t)^n\, dW(t),
\end{equation}
with real constants $n$ and $c$.
Its formal solution is 
\begin{equation}
v(t) =  \left\{\begin{array}{cl} \left[ v(0)^{1-n} + c(1-n)\,\xi(t) \right]^{\frac{1}{1-n}} & \text{if $n\neq 1$,}\\
v(0)\,\exp\left[\,c\xi(t)\,\right] & \text{if $n=1$.}
\end{array}\right.
\end{equation}
where again $\xi(t) = \int_{s=0}^{\,t} dW(s)$
is a Gaussian random variable of variance $t$, mean zero.
One finds that:
\begin{equation}
\langle v^m \rangle_{\rm s} \left\{\begin{array}{cccc}\text{ converges } & \forall m &\text{ and } \forall t &\text{ for } n\le 1 \\
\multicolumn{3}{c}{\text{does not converge}} &\text{ for } n>1 \end{array}\right.
\end{equation}
and
\begin{equation}
\fl\langle \exp(v^m) \rangle_{\rm s} \left\{\begin{array}{cccc}
\text{ converges } & \forall m < 2(1-n) &\text{ and } \forall t &\text{ for } n < 1 \\
\text{ converges } & \text{ for } m = 2(1-n) &\text{ and } \text{ for } t<\frac{1}{2c(1-n)^2} &\text{ for } n < 1 \\
\multicolumn{3}{c}{\text{does not converge}} &\text{ for } n\ge1 \end{array}\right.
\end{equation}

One can see from the above that, barring special favourable circumstances\footnote[1]{e.g. a topological barrier which prevents any trajectories reaching $|v|\to\infty$, given the right initial conditions.}, the following relationships will hold (the notation is as in (\ref{genericeqn})):

A) If observable averages involve only expressions polynomial in the variables $v$, then schemes for which noise terms grow faster than \emph{linearly} as $|v|\to\infty$ will be divergent. That is those where 
\begin{equation}\lim_{|v|\to\infty}\re{B_v(\bo{v})/v}\label{nwa}\end{equation}
 is unbounded for any $v$. For example $B_v=cv^n$, where $n>1$.

B) If observable averages involve expressions exponential in variables $v$, then schemes for which noise terms grow faster than $\propto \sqrt{|v|}$ as $|v|\to\infty$ will be divergent. That is those where 
\begin{equation}\lim_{|v|\to\infty}|B_v(\bo{v})/\sqrt{v}\,|\neq 0.\label{nwb}\end{equation}
The special case when the limit (\ref{nwb}) is finite is also divergent, but only after some initial time period.

\subsection{Single mode: diffusion gauge}
How do the gauged methods compare to these divergence symptoms for the single-mode system? 
The diffusion-gauged method of Section~\ref{DIFF} has a finite formal solution (31) at all times. 
For observables of finite order in $\dagop{a}$ and $\op{a}$, only polynomial expressions of $\alpha$ and $\beta$ 
need be averaged. Such expressions  scale as $\propto\exp[ \text{factors} \times \xi_j(t)]$, and thus their stochastic averages are equivalent to integrals like $\int P(\xi) \exp[\text{factors}\times\xi]\, d\xi$ which are convergent due to the Gaussian form of $P(\xi)$.

\subsection{Single mode: drift gauge}
The equations (\ref{1modegaequations}) with the drift gauge (\ref{driftgauge1}) can be formally solved:
\begin{eqnarray}
\log n(t) &=& \log n(0) -\gamma t - e^{-g''}\sqrt{\kappa}\left[\xi^-(t)-i\xi^+(t)\right]\\
\log \Omega(t) &=& -e^{-g''}\sqrt{\kappa}\,\int_{s=0}^{\,t}\im{n(t)}\left[dW_1(s)-idW_2(s)\right].
\end{eqnarray}
(taking $\Omega(0)=1$). From (\ref{moments}), observables require the averaging of quantities like 
$\Omega f(\alpha,\beta)$, which include the factor $|\Omega(t)|$. This takes the form
\begin{equation}
\fl|\Omega(t)| = \exp\left\{ -\sqrt{\kappa}e^{-g''}\int_{s=0}^{\,t} 
|n(0)|e^{-\gamma s}e^{-\sqrt{\kappa}e^{-g''}\xi^-(s)}\sin\left[\angle n(0)+\sqrt{\kappa}\xi^+(s)\right]\,dW^+(s) \right\},
\end{equation}
where $dW^+=(dW_1+dW_2)/\sqrt{2}$. 

For large negative $\xi^-(s)$, the gaussian drop-off of $P(\xi^-(s))$ is insufficient by itself to directly prevent the divergence of this factor. However, the situation is more subtle than in the simple model equations, due to the presence of an oscillatory stochastic
term in the integral. Hence the convergence of the observable averages is still an open question for the drift-gauged method, despite the absence of movable singularities.

Lack of convergence has been shown for the ``simple coherent'' stochastic wavefunction method\cite{CCconvergence}, which  has been discussed in Section~\ref{CH7ComparisonCCD}.
  
\subsection{Many-modes}
Regarding movable singularities in the many-mode equations (11), let us first compare to the movable singularity condition (\ref{mvsingc}).
One sees that: 

1) The drift terms dependent on $\omega_{\bo{n}\bo{n}}$ and $\gamma_{\bo{n}}$
as well as the noise terms dependent on $\kappa$ lead to only exponential
growth in the moduli of $\alpha_{\bo{n}}$ and $\beta_{\bo{n}}$,
and so can not cause movable singularities. 

2) The mode-mixing terms dependent on $\omega_{\bo{n}\bo{m}}$, where
$\bo{n}\neq\bo{m}$, lead to drift of the form \begin{eqnarray}
d|\alpha_{\bo{n}}| & = & |\alpha_{\bo{m}}||\omega_{\bo{n}\bo{m}}|\sin\left(\angle\omega_{\bo{n}\bo{m}}+\angle\alpha_{\bo{m}}-\angle\alpha_{\bo{n}}\right)dt+\dots\nonumber \\
 & \leq & |\alpha_{\bo{m}}||\omega_{\bo{n}\bo{m}}|dt+\dots.\end{eqnarray}
 and of similar form for $|\beta_{\bo{n}}|$. These terms lead to
behaviour like a linear matrix differential equation for the radial
evolution of the coherent state amplitudes. Such equations do not
diverge in finite time, their solutions being finite linear combinations
of exponentials, hence these terms by themselves can not lead to movable
singularities either. 

3) Only the nonlinear drift terms can cause movable singularities.
With no drift gauge ($\mc{G}_{k}=0$) as in the positive P or diffusion-gauge-only
equations, these lead (when dominant) to evolution of the form \begin{eqnarray}\eqalign{\label{badterms}
d|\alpha_{\bo{n}}| & =  \kappa|\alpha_{\bo{n}}|^{2}|\beta_{\bo{n}}|\sin(\angle\alpha_{\bo{n}}+\angle\beta_{\bo{n}})+\dots\\
d|\beta_{\bo{n}}| & =-\kappa|\beta_{\bo{n}}|^{2}|\alpha_{\bo{n}}|\sin(\angle\alpha_{\bo{n}}+\angle\beta_{\bo{n}})+\dots\,\,.}\end{eqnarray}
 \textit{\emph{These}} \textit{can} cause either $|\alpha_{\bo{n}}|$
or $|\beta_{\bo{n}}|$ to diverge at finite time, for some trajectories.
For example, if $|\beta_{\bo{n}}|\sin(\angle\alpha_{\bo{n}}+\angle\beta_{\bo{n}})=K$
conspire to be approximately constant and nonzero over the relevant
timescale, then \begin{equation}
|\alpha_{\bo{n}}(t)|\approx\frac{|\alpha_{\bo{n}}(t_{0})|}{1-\kappa K|\alpha_{\bo{n}}(t_{0})|(t-t_{0})},\end{equation}
 which diverges at time $t_{\mathrm{div}}=t_{0}+1/\kappa K|\alpha_{\bo{n}}(t_{0})|$.
Of course the precise condition {}``$K$ is constant over the time
$t_{0}$ to $t_{\mathrm{div}}$'' will only occur for a set of trajectories
of measure zero, but this is typical of movable singularities. 

4) The equations that use the drift gauges (\ref{driftgauge}) do
not contain the terms (\ref{badterms}), and no movable singularities occur.

As for noise-weight divergences, the situation appears to be similar to the single-mode cases. 
The formal solution for the weight is 
\begin{equation}
\fl\qquad\Omega(t) = \exp\left\{-i\sqrt{i\kappa}\int_{s=0}^{\,t}e^{-g''(s)}\sum_{\bo{n}} \im{n_{\bo{n}}(s)} \times\text{noise increments}(s)\right\},
\end{equation}
and contains an exponent containing exponentials of gaussian random variables $\xi^-_k(s)$ as part of the $n_{\bo{n}}(s)$.
This is 
indicative of possible divergences.

\subsection{The ``what happened to the divergence?'' puzzle}
It appears that the observable averages using the drift gauge (\ref{driftgauge1}) have the potential to be non-convergent even in the single mode case. Why then are the numerical simulations of Section~\ref{1MODE} well behaved for such a long time, showing no sign of bias? Similarly, no systematic error was seen in simulations using the ``simple coherent'' stochastic wavefunction scheme\cite{carusottodynamix}, despite the subsequent proof of its divergence in Ref.~\cite{CCconvergence}. 

The detailed investigation of this is beyond the scope of this paper. The simplest
explanation is simply that the distribution tails are sufficiently convergent
to eliminate boundary terms, while still having a large (perhaps infinite)  variance
in some observables. It is  possible that 
the appearance of large statistical uncertainty masks any systematic errors that may occur in 
the $\mc{S}\to\infty$ limit. This is plausible because long distribution tails certainly give rise to large phase-space excursions and thus huge statistical uncertainty, whether or not systematic biases in the limit $\mc{S}\to\infty$ are present. Another reason for this lack of bias may be some type of special symmetry properties in the equations or the observables calculated.

\subsection{Summary}
Divergences of the moving singularity type may be present in many-mode (but not single-mode) simulations using the diffusion-gauge-only method of Section~\ref{DIFF}, while divergences of the noise-weight type may be present in simulations with the drift-gauged method described in Section~\ref{DRIFT}. 
However, no bias of any kind was seen in the single-mode simulations (or the two-mode simulations, as shall be seen below). 

Hence, the simulations appear to give correct results, but numerical indicators such as spiking and ensemble-size dependence should be rigorously monitored in all calculations.

\section{Two coupled modes}

\label{2MODE}

We now look at the behaviour of a two-mode system, to investigate
how the statistical behaviour seen for the single mode is affected
once coupling between modes is present. This is a simple enough system
that investigation of several examples gives meaningful insight into
the general situation.

\subsection{The model}

\label{CH8Model}

The system consists of two orthogonal modes labeled $1$ and $2$
with inter-particle interactions in each mode, and Rabi coupling between
them. No damping will be considered, for simplicity. The coupling
frequency will be restricted to be real. Time units are chosen so
that the nonlinear interaction frequency is $\kappa=2$, and a transformation
is made to an interaction picture in which the linear mode self-energies
$\hbar\omega_{jj}\op{n}_{j}$ are moved into the Heisenberg evolution
of operators. The interaction picture Hamiltonian then is \begin{equation}
\op{H}=\hbar\omega_{12}\left[\dagop{a}_{1}\op{a}_{2}+\dagop{a}_{2}\op{a}_{1}\right]+\hbar\sum_{j=1}^{2}\dagop{a}_{j}{}^{2}\op{a}_{j}^{2}.\label{2H}\end{equation}
 The Rabi frequency is $\omega_{12}$ (in scaled time units). Stochastic
equations are as in \eqref{itoequations}. Physically this model can
represent, for example, two internal boson states coupled by an EM
field, or two trapped condensates spatially separated by a barrier.
This approximation has been widely used to investigate the quantum
behavior of BECs in two-state systems\cite{2modes}. 

Let us consider two kinds of initial conditions which broadly represent
the two kinds of situations generically occurring in all many-mode
simulations: 

\begin{enumerate}
\item Case 1: Coupling between modes of widely differing occupation. 
\item Case 2: Coupling between modes of similar occupation. 
\end{enumerate}
In a many-mode calculation, adjacent modes typically behave like case
2, since if a field model is well resolved by the lattice, then physical
properties (e.g. density, and hence mode occupation) should not change
much over the distance between neighboring lattice points. Long distance
coupling will tend to behave like case 1.

\subsection{Case 1: Coupling to a vacuum mode}

\label{CH8Case1}

The system starts initially with a coherent state of mean particle
number $n_{0}$ in mode 1, and vacuum in mode 2. In all simulations
of this case, the inter-mode coupling strength was taken to be $\omega_{12}=5$,
but the mean particle number $n_{0}$ (conserved in time) was varied. 

At low particle number, the Rabi oscillations dominate the Hamiltonian,
and particles oscillate between the modes, without much phase collapse.
At high particle number $n_{0}$, on the other hand, phase collapse
dominates mode 1, suppressing also the coherent transfer of particles
to mode 2. 

The particular values chosen to simulate were \begin{equation}
n_{0}=\{1,17,200,1500,10^{4}\},\end{equation}
 Simulation times were assessed using the calculated uncertainties
in the two observables: 1) The fraction of particles in the (initially
empty) mode 2: \begin{equation}
p_{2}=\frac{\langle\op{n}_{2}\rangle}{\ba{N}},\end{equation}
 where $\op{n}_{j}=\dagop{a}_{j}\op{a}_{j}$, and 2) the local normalized
second order correlation functions: \begin{equation}
g_{j}^{(2)}(t,t)=\frac{\langle:\op{n}_{j}(t):\rangle}{\langle\op{n}_{j}(t)\rangle}=\frac{\langle\op{n}_{j}^{2}\rangle}{\langle\op{n}_{j}\rangle^{2}}-\frac{1}{\langle\op{n}_{j}\rangle}.\label{g2def}\end{equation}

The second order correlations quantify the amount of (instantaneous)
bunching/antibunching in the boson field. These are unity for coherent
states, two for thermal fields, and $1-1/n$ for Fock number states
of $n$ particles. Large values $g^{(2)}>2$ occur e.g. for quantum
superpositions of vacuum and Fock number states with two or more particles
where the average particle number is small. For example in the state
$|\psi\rangle=\sin\theta|0\rangle+\cos\theta|n\rangle$, $g^{(2)}=(1-\frac{1}{n})/\cos^{2}\theta$. 

The drift and diffusion gauge scheme using \eqref{driftgauge} and
\eqref{giiga} was considered. Comparison was also made to the positive
P ($g''_{j}=\mc{G}_{k}$=0), and to the special case of $t_{\textrm{opt}}=0$ in the 
diffusion gauge which then is \begin{equation}
g''_{\bo{n}}=\frac{1}{6}\log\left[1+4n''_{\bo{n}}(t)^{2}\right].\label{giiga0}\end{equation}
 This may become nonzero after spread in the $n''_{j}$ from the
initial $n''_{j}=0$ occurs. In each run, $\mc{S}=2\times10^{5}$
trajectories were used, and useful simulation precision taken to occur at such a time $t_{\rm sim}$ 
when 10\% or smaller relative uncertainty in an observable could be
obtained using $10^{6}$ trajectories. 

\begin{figure}
\center{\includegraphics[width=\columnwidth]{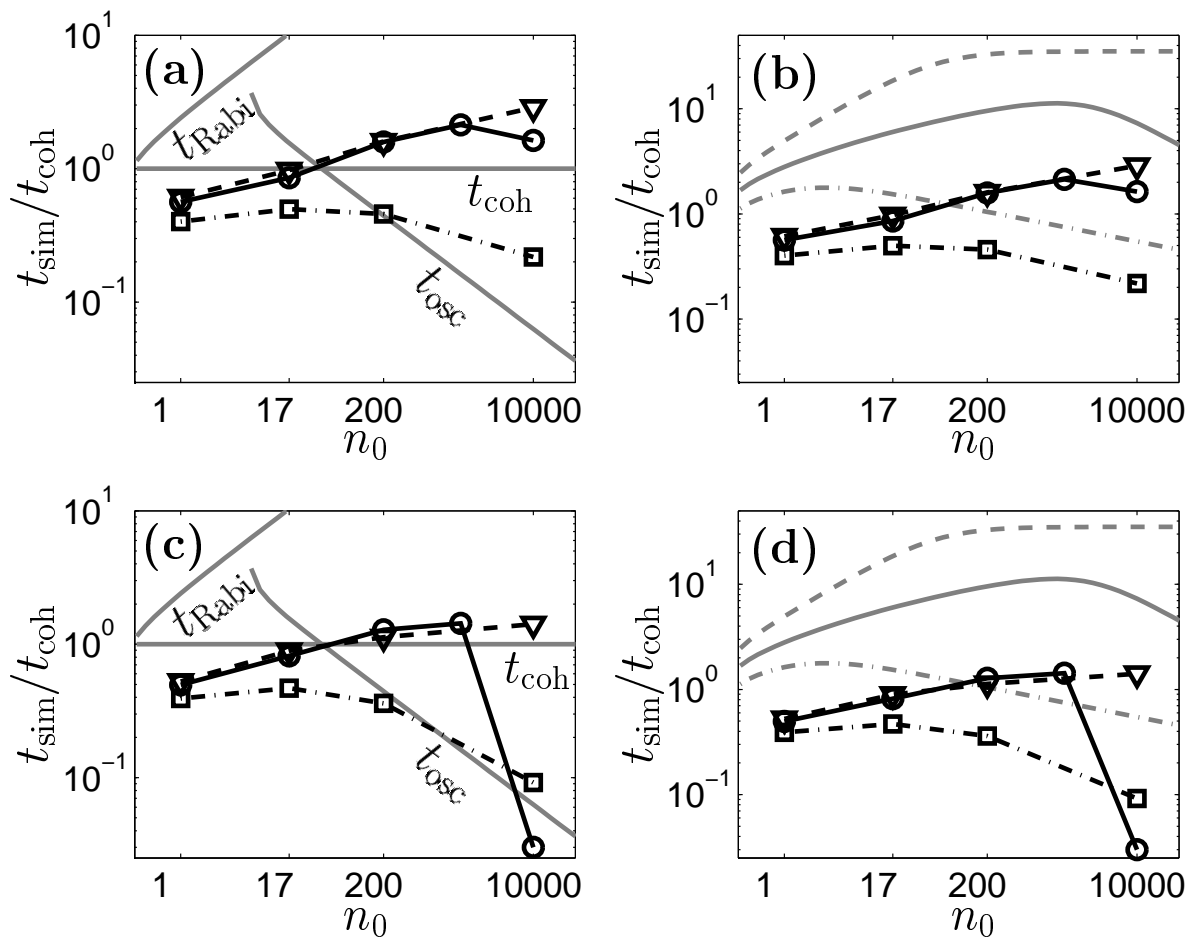}}\vspace*{-10pt}
\caption{\label{FIGtsimcase1} Useful simulation times $t_{\textrm{sim}}$
for a mode coupled to vacuum as in Section~\ref{CH8Case1}. Calculated
simulation times are shown as data points, with the symbols denoting
gauge used. {}``$\Box$'': positive P; {}``$\bigcirc$'': drift
gauge \eqref{driftgauge} and diffusion gauge \eqref{giiga0}; {}``$\bigtriangledown$'':
drift and diffusion gauges \eqref{driftgauge} and \eqref{giiga}
with best target time parameters: $\omega_{12}t_{\textrm{opt}}=\{2.5,0.5,0.5,0.075\}$
for $n_{0}=\{1,17,200,10^{4}\}$, respectively. Dependence of $t_{\textrm{sim}}$
on $t_{\textrm{opt}}$ can be found in\cite{inprep}, Figure~8.6.
Subplots \textbf{(a)} and \textbf{(b)} show simulation times based
on estimates of the observable $p_{2}$, while \textbf{(c)} and \textbf{(d)}
times based on $g_{2}^{(2)}$. Subplots \textbf{(a)} and \textbf{(c)}
compare to physical time scales, including Rabi oscillation period
$t_{\textrm{Rabi}}=2\pi/\omega_{12}$, while subplots \textbf{(b)}
and \textbf{(d)} compare to expected simulation times for a single
mode using the empirical fits of Table~\ref{TABLEtsimfit} and \eqref{fitga0}.
The expected $t_{\textrm{sim}}$ are plotted as light lines. Dotted:
positive P; solid: drift and diffusion gauges with $t_{\textrm{opt}}=0$;
dashed: with optimum $t_{\textrm{opt}}$ choice. }
\end{figure}

\begin{figure}
\center{\includegraphics[width=\columnwidth]{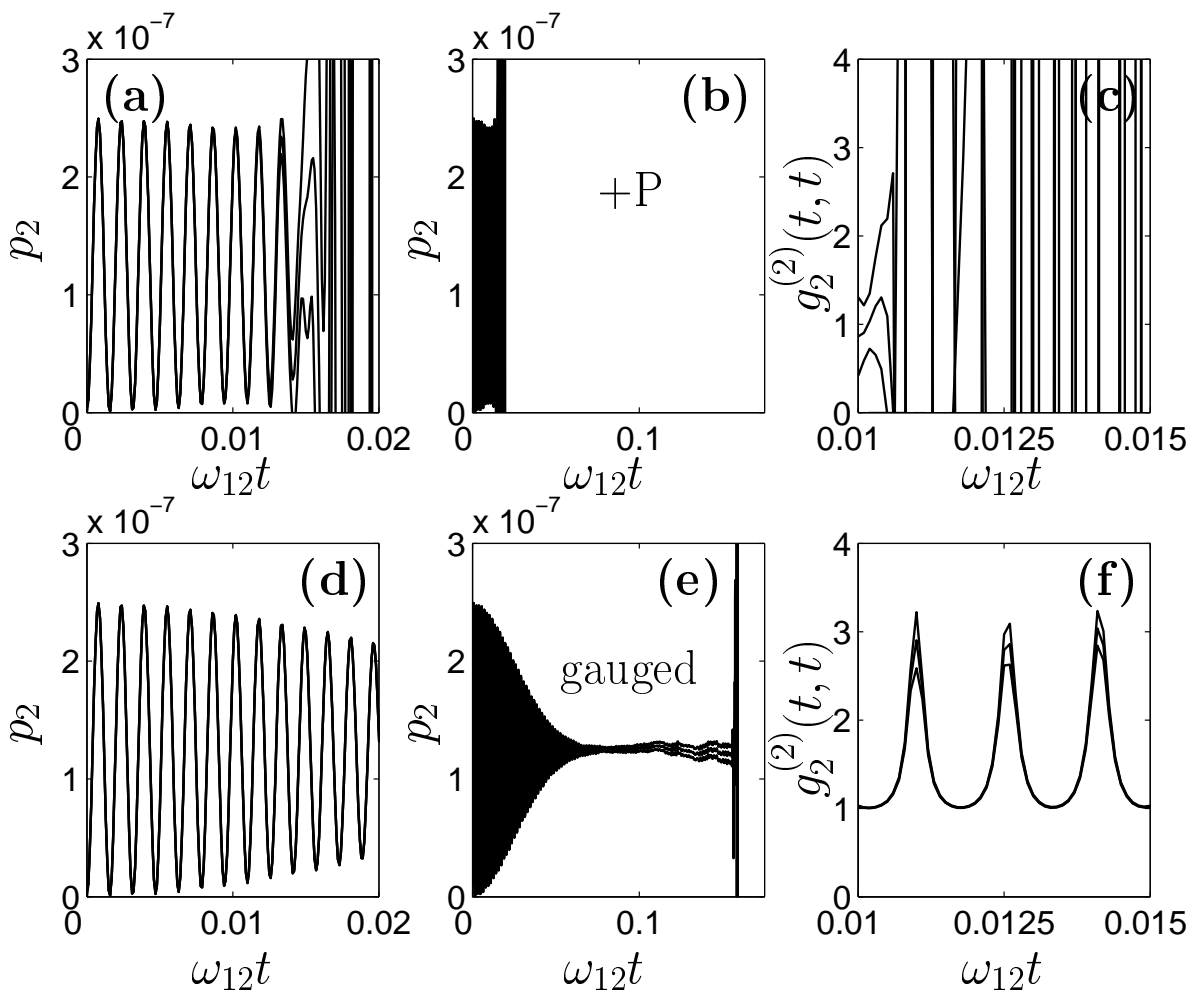}}
\caption{\label{FIGn1e4case1} Coupling to vacuum mode as in Section~\ref{CH8Case1},
$n_{0}=10^{4}$. Subfigures \textbf{(a)}--\textbf{(c)} show results
with positive P simulations, whereas \textbf{(d)}-- \textbf{(f)} show
results with combined drift gauge (\ref{driftgauge}) and diffusion
gauge (\ref{giiga}) using target time $\omega_{12}t_{\text{opt}}=0.1$.
Triple lines indicate mean and error bars. }
\end{figure}

In Figure~\ref{FIGtsimcase1} simulation times are compared to physical
timescales and expected values based on single-mode expressions from
Table~\ref{TABLEtsimfit}. An example simulation is shown in Figure~\ref{FIGn1e4case1}.
Note that for gauged simulations using \eqref{driftgauge} and \eqref{giiga0},
single-mode simulations led to the $t_{\textrm{sim}}$ empirical fitting
parameters \begin{equation}
\{ c_{0},\dots,c_{4}\}=\{1,800\pm260,3.6\,{}_{-2.3}^{+\infty},1.2\pm0.2,0.42\pm0.03\}.\label{fitga0}\end{equation}
 One point to note is that $t_{\textrm{sim}}$ was based on the moment
when relative error in a quantity $g_{j}$ or $p_{j}$ was \textit{first}
found to be too large. In calculations of $g^{(2)}$, uncertainties
are much greater when $g^{(2)}(0,t)$ peaks --- see e.g. Figure~\ref{FIGn1e4case1}\textbf{(f)}.
Good accuracy can often be obtained between peaks for much longer
times than shown in Figure~\ref{FIGtsimcase1}\textbf{(c)} or \textbf{(d)},
up to about the same simulation time as worked out based on $p_{2}$.
This is especially evident for $n_{0}=10^{4}$. 

\subsection{Case 2: coherent mixing of two identical modes}

\label{CH8Case2}

The system starts initially with identical coherent states of mean
particle number $n_{0}$ in both modes. The two-mode state is separable.
This time simulations were carried out with constant particle number
$n_{0}=100$, but the coupling frequency $\omega_{12}$ was varied. 

At low frequency $\omega_{12}\ll n_{0}=100$, phase collapse local in
each mode dominates, and phase oscillations in each mode occur with
period $t_{\textrm{osc}}=\pi/n_{0}$, while at high frequency $\omega_{12}\gg n_{0}=100$,
the inter-mode coupling dominates and phase oscillations for each
mode occur with period $t_{\textrm{Rabi}}=2\pi/\omega_{12}$. One
expects that for weak coupling the two modes should behave largely
as two independent single modes of Section~\ref{1MODE}. 

The particular values chosen to simulate were \begin{equation}
\omega_{12}=\{5000,500,50,5,0.5,0.05,0.005,0.0005\}.\end{equation}
 The simulation time $t_{\textrm{sim}}$ was assessed here based on
10\% uncertainty in $|G^{(1)}(0,t)|$ (for either mode), as in the
single-mode case. There were $\mc{S}=10^{4}$ trajectories per simulation. 

\begin{figure}
\center{\includegraphics[width=400pt]{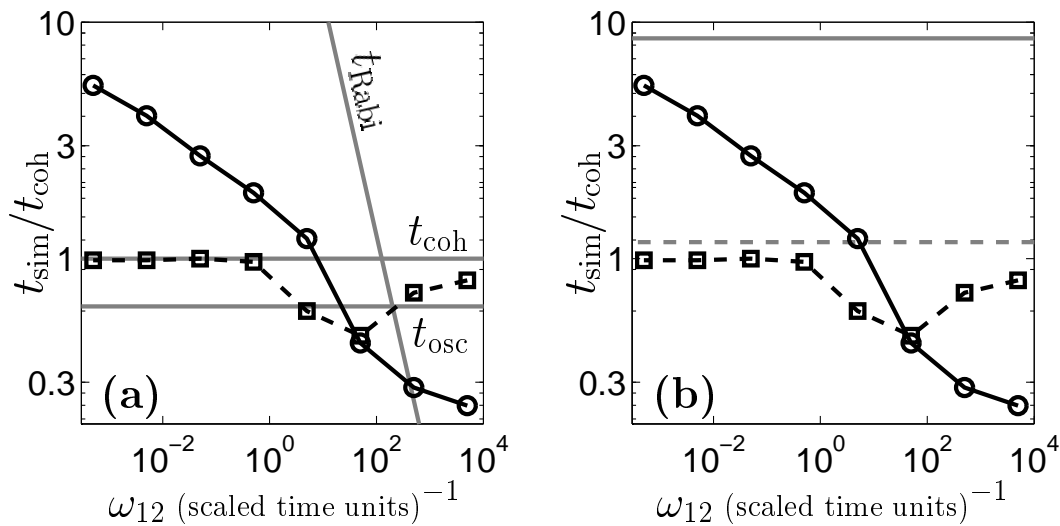}}
\caption{\label{FIGtsimcase2} Useful simulation times $t_{\textrm{sim}}$
for a two identical modes undergoing coherent mixing as in Section~\ref{CH8Case2}.
Calculated simulation times are shown as data points, with the symbols
denoting gauge used. {}``$\Box$'': positive P; {}``$\bigcirc$'':
drift gauge \eqref{driftgauge} and diffusion gauge \eqref{giiga0}.
Subplot \textbf{(b)} compares to expected simulation times for a single
mode using the empirical fits of Table~\ref{TABLEtsimfit} and \eqref{fitga0}.
These expected $t_{\textrm{sim}}$ are plotted as light lines: dashed:
positive P; solid: drift gauged. }
\end{figure}

\begin{figure}
\center{\includegraphics[width=\columnwidth]{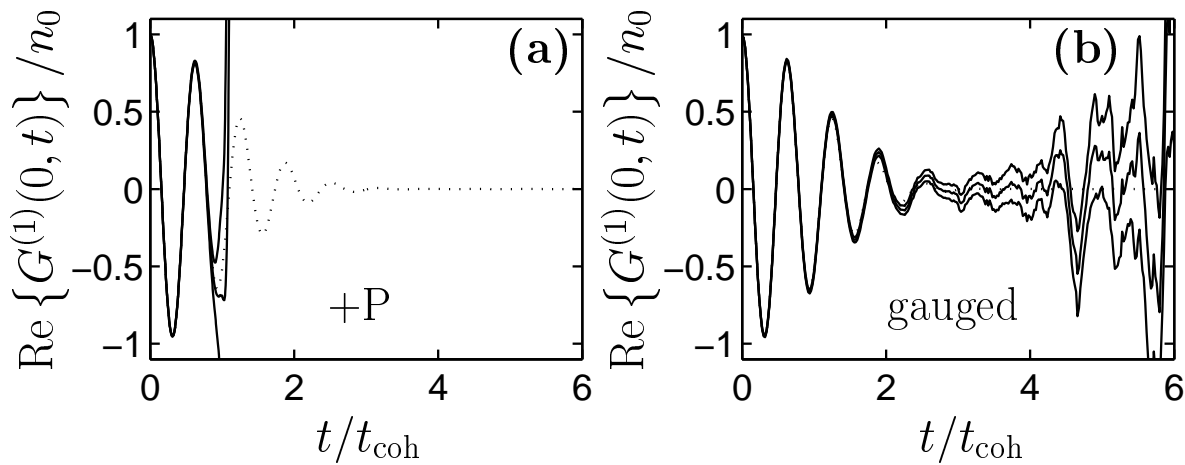}}\vspace*{-10pt}
\caption{\label{FIGcase2} Mixing of two identical modes as in Section~\ref{CH8Case2},
$\omega_{12}=0.0005$. Subplot \textbf{(a)} shows results with a positive
P simulation, whereas \textbf{(b)} shows results with combined drift
gauge (\ref{driftgauge}) and diffusion gauge (\ref{giiga0}). Triple
lines indicate mean and error bars. }
\end{figure}

In Figure~\ref{FIGtsimcase2} simulation times are compared to physical
timescales and expected values based on single-mode expressions using
Table~\ref{TABLEtsimfit}. An example simulation is shown in Figure~\ref{FIGcase2}. 

Note that the $t_{\textrm{opt}}=0$ gauge \eqref{giiga0} performed
better than any $t_{\textrm{opt}}>0$ gauge for all $\omega_{12}$
values tried here. The dependence of $t_{\textrm{sim}}$ on $t_{\textrm{opt}}$
is shown in Figure~8.8 of Ref.~\cite{inprep}. 

\subsection{Analysis}

\label{CH8Analysis}

The above examples have not by any means been a comprehensive assessment
of gauge performance for general cases of the model \eqref{2H}, since
only a few parameter regimes have been explored. Still, several aspects
of the situation when modes are coupled have been seen: 

\begin{itemize}
\item Broadly speaking, when local scattering within a mode dominates over
the coupling between modes, the response of the system to local gauges
is similar to what was seen for single modes. Large extensions of $t_{\textrm{sim}}$
are found relative to the positive P method to well beyond $t_{\textrm{coh}}$
when $n\gg1$. Little improvement is found when $n\lesssim\order(1)$.
In this strong scattering regime, one doesn't need much inter-mode
coupling $\omega_{12}$ to reduce the simulation time in absolute
terms by a factor $\order(2-10)$, although $t_{\textrm{sim}}>t_{\textrm{coh}}$
is still obtained. 
\item When the inter-mode coupling dominates, the local diffusion gauges
do not appear to be useful. They actually reduce simulation time as
compared to the positive P method, although several Rabi oscillation
periods can always be simulated. In the Case 2 simulations, the transition
between the strong and weak coupling behavior appears to be at around
\begin{equation}
\omega_{12}\approx\order(\kappa n),\end{equation}
 which is the point at which the expectation values of the coupling
and two-body scattering energies in the Hamiltonian are approximately
equal. This implies, as noted previously, that nonlocal gauges are
likely to be more useful in these cases.
\item The beneficial effect of choosing $t_{\textrm{opt}}>0$ seen for a
single mode (see e.g. Figure~\ref{FIGtopt}) appears to be suppressed
at intermediate mode occupations $n_{0}\lesssim\order(200)$. However,
the $t_{\textrm{opt}}=0$ diffusion gauge \eqref{giiga0} has a marked
beneficial effect even when two body scattering is significant. This
parameter range is for $\omega_{12}\ll n\kappa$, and $\order(1)\lesssim n_{0}\lesssim\order(10^{3})$.
Simulation times obtained are smaller by about a factor of $\order(2)$
than those given for a single mode with the same gauge. At higher
occupations, the benefit gained with \eqref{giiga0} abates but nonzero
$t_{\textrm{opt}}$ values appear to become useful again, and continue
to provide strong improvements over positive P simulations. (See,
e.g. Figure~\ref{FIGtsimcase1}\textbf{(d)}.) The gauge \eqref{giiga0}
is convenient also because there is no \textit{a priori} parameter
$t_{\textrm{opt}}$. 
\end{itemize}
At the level of the stochastic equations, the evolution of $d\alpha_{j}$
can gather randomness from three sources 

\begin{enumerate}
\item Directly from the local noise term $\propto\alpha_{j}\sqrt{\kappa}\, dW_{j}$
\item Indirectly from the local nonlinear term $\propto\kappa\alpha_{j}n_{j}$
that can amplify variation in the local noise term. 
\item From the other mode through the coupling term $\propto\omega_{12}\alpha_{\neg j}$. 
\end{enumerate}
The drift gauges \eqref{driftgauge} neutralize source 2. The diffusion
gauges \eqref{gii}, \eqref{giiga} or \eqref{giiga0} suppresses
the direct noise source 1. No local gauge is good at suppressing the
third source of randomness, however, because these fluctuations are
largely independent of any processes occurring in mode $j$. What
happens is that even small randomness in one mode feeds into the other,
can become amplified, and fed back again. Combating such effects would require
a nonlocal gauge.

Lastly, significantly more detail on two-mode simulations, including
consideration of some other gauges, and the relationship between $t_{\textrm{opt}}$
and $t_{\textrm{sim}}$ can be found in Chapter 8 of Ref.~\cite{inprep}.

\section{Many-mode example: uniform gas}

\label{MMODE}

We revisit the uniform gas system simulated in the 
companion paper\cite{paperA}. This consists of a uniform one-dimensional
gas of bosons with density $\rho$ and inter-particle \mbox{$s$-wave}
scattering length $a_{s}$. The lattice is chosen with a spacing $\Delta{\textrm{x}}\gg a_{s}$
so that inter-particle interactions are effectively local at each
lattice point, and the lattice Hamiltonian \eqref{Hamiltonian} applies.
Periodic boundary conditions are assumed. 

The initial state is taken to be a coherent wavefunction, which is
a stationary state of the ideal gas with no inter-particle interactions
(i.e. $a_{s}=0$). Subsequent evolution is with constant $a_{s}>0$,
so that there is a disturbance at $t=0$ when inter-particle interactions
are rapidly turned on. Physically, this can correspond to the disturbance
created in a BEC by rapidly increasing the scattering length at $t\approx0$
by e.g. tuning the external magnetic field near a Feshbach resonance\cite{Donley,Bosonic-Cs-Rb-Na}. 
More discussion of the physics can be found in the companion paper\cite{paperA}.

A brief set of exploratory simulations were made with the diffusion
gauge \eqref{gii} (no drift gauge) to investigate whether simulation
time can be extended. The case of $\rho=100/\xi^{\textrm{heal}}$
was picked, and target time $t_{\textrm{opt}}$ was varied for the
two cases $\Delta{\textrm{x}}=\xi^{\textrm{heal}}/2$ and $\Delta{\textrm{x}}=10\xi^{\textrm{heal}}$.
Note that in both cases the occupation per mode $\ba{n}=\rho\Delta{\textrm{x}}$
is $\gg1$. Simulations were with $M=250$ and $M=50$ lattice points
(meaning $12500$ and $5\times10^{4}$ particles on average), respectively,
and $\mc{S}=10^{4}$ trajectories in both cases. 

\begin{figure}
\center{\includegraphics[width=\columnwidth]{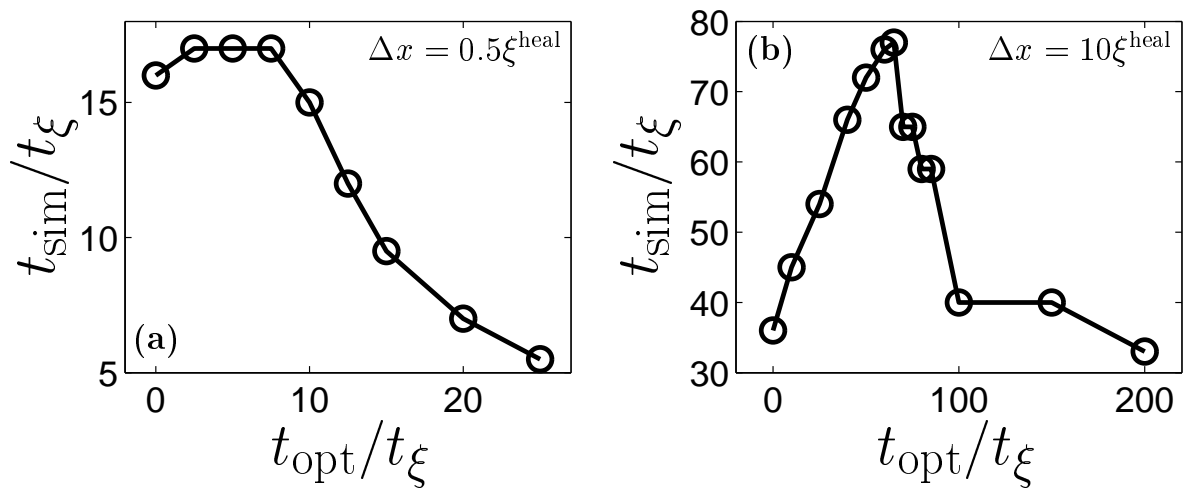}}\vspace*{-10pt}
\caption{\label{FIGMtopt} Simulation times as a function of the target time
$t_{\textrm{opt}}$ when using the diffusion gauge \eqref{gii}, and
$\mc{G}_{k}=0$. Standard positive P method when $t_{\textrm{opt}}=0$.
All simulations are of a $\rho=100/\xi^{\textrm{heal}}$ gas, but
with differing lattice spacing $\Delta x$. $\mc{S}=10^{4}$ trajectories,
$M=50$. }
\end{figure}

Simulation times obtained are shown in Figure~\ref{FIGMtopt}, and
correspond to the time of first spiking seen in estimates of the second-order
spatial correlation function \begin{equation}
\ba{g}^{(2)}(\bo{x}_{\bo{n}})=\frac{1}{M}\sum_{\bo{m}}\frac{\langle\dagop{a}_{\bo{m}}\dagop{a}_{\bo{m+n}}\op{a}_{\bo{m}}\op{a}_{\bo{m+n}}\rangle}{\langle\dagop{a}_{\bo{m}}\op{a}_{\bo{m}}\rangle\langle\dagop{a}_{\bo{m+n}}\op{a}_{\bo{m+n}}\rangle}.\end{equation}
 The time scale used is \begin{equation}
t_{\xi}=\frac{m(\xi^{\textrm{heal}})^{2}}{\hbar}=\frac{\hbar}{2\rho g}\label{txidef}\end{equation}
 (the {}``healing time''), which is approximately the time needed
for the short-distance $\order(\xi^{\textrm{heal}})$ inter-atomic
correlations to equilibrate after the disturbance. See e.g. Ref.~\cite{paperA}. 

For comparison, gauged simulations of isolated lattice points with
the same occupations $\ba{n}=\rho\Delta{\textrm{x}}$ gave much bigger
improvements in $t_{\textrm{sim}}$. Using \eqref{timefit} and diffusion
gauge \eqref{gii} data from Table~\ref{TABLEtsimfit}, we expect
$t_{\textrm{sim}}\approx140t_{\xi}$ and $t_{\textrm{sim}}\approx660t_{\xi}$
for $\Delta{\textrm{x}}=2\xi^{\textrm{heal}}$ and $10\xi^{\textrm{heal}}$,
respectively. This is to be contrasted with the predictions of $t_{\textrm{sim}}\approx15t_{\xi}$
and $t_{\textrm{sim}}\approx49t_{\xi}$, respectively, for the standard
positive P method, which are also an accurate reflection of $t_{\textrm{sim}}$
in the present many-mode case. This is consistent with what was seen
in Ref.~\cite{paperA}. 

Clearly, the local gauges extend simulation time in these large uniform density systems, although the improvement is relatively smaller than for the single
mode. Also, the numerical results are consistent with the analysis in Section~\ref{MODEDIFF}
which indicated in  \eqref{dxheal} that simulation time 
time improvements would occur once \mbox{$\Delta{\textrm{x}}\gtrsim\order(\xi^{\textrm{heal}})$}
. Lastly, it is noteworthy
that the dependence of $t_{\textrm{sim}}$ on $t_{\textrm{opt}}$
shows qualitative similarities to the single-mode case of Figure~\ref{FIGtoptnopt}.

\section{Conclusions}

\label{CONCLUSIONS}

The positive P representation method is capable of simulating many-body
quantum dynamics of interacting Bose gases from first principles\cite{evaporativecooling,samperrorinbec,paperA,inprep}.
A limiting factor is that precision is lost after a certain time $t_{\textrm{sim}}$
due to sampling error, and this may be accompanied by systematic boundary term errors. This time can be short when some modes of a
many-mode boson system are occupied by many particles. In that case,
precision is lost before these highly occupied modes lose coherence due to phase diffusion\cite{paperA}. 

Using the related gauge P representation, local diffusion and drift
gauges have been developed that greatly improve the useful simulation
times at high occupation, most strongly for single-mode cases. The resulting
simulation times have been investigated in some detail, and substantial
improvement has been demonstrated in 1, 2, and many-mode ($M=50$ and $M=250$)
simulations. For many of the single- and double-mode cases considered,
full decoherence can be achieved while still retaining good precision. 

Two gauge choices were proposed as being the most advantageous: 

\begin{enumerate}
\item Diffusion gauge \eqref{gii} only. 
\item Drift \eqref{driftgauge} and diffusion gauge \eqref{giiga}. 
\end{enumerate}
The latter appears less broadly applicable because weight fluctuations
accumulate from all modes, leading to a decrease in simulation time
with increased system size, and because of some doubt over its convergence properties.
 However, this latter method can lead to
longer simulation times when there is one, or only several, dominant modes.

Considerations of convergence in Section~\ref{BT} have highlighted 
two routes that may lead to divergences: moving singularities and noise-weight considerations. Nevertheless, 
no sign of any bias is seen, an interesting situation which warrants further investigation. 

Generally speaking, the local-gauge methods considered here give much
better performance when the inter-particle scattering local to each
mode is a stronger process than inter-mode coupling due to kinetic
evolution and/or external potentials. This is the case if sufficiently
coarse lattices are used. 

Finally, the investigation of local gauges carried out here may also
be useful in developing more robust nonlocal gauges. These might lengthen
simulation times for lattices denser than the healing length, beyond
what is possible with either the standard positive P method,
or the local stochastic gauges.


\section*{References}
\begin{thebibliography}{10}
\bibitem{Wig-Wigner}E.~P. Wigner, Phys.~Rev. \textbf{40}, 749 (1932). 
\bibitem{Hus-Q}K.~Husimi, Proc.~Phys.~Math.~Soc.~Jpn. \textbf{22}, 264 (1940). 
\bibitem{Gla-P}R.~J.~Glauber, Phys.~Rev. \textbf{131}, 2766 (1963); E.~C.~G.~Sudarshan,
Phys.~Rev.~Lett. \textbf{10}, 277 (1963).
\bibitem{positiveP1}S.~Chaturvedi, P.~D.~Drummond, and D.~F.~Walls, J.~Phys.~A
\textbf{10}, L187-192 (1977). 
\bibitem{positiveP2}P.~D.~Drummond and C.~W.~Gardiner, J.~Phys.~A \textbf{13}, 2353
(1980). 
\bibitem{removalofboundaryterms}P.~Deuar and P.~D.~Drummond, Phys.~Rev.~A \textbf{66}, 033812
(2002). 
\bibitem{stochasticgauges}P.~D.~Drummond and P.~Deuar, J.~Opt.~B-Quant.~and Semiclass.~Opt.
\textbf{5}, S281 (2003). 
\bibitem{carusottodynamix}I.~Carusotto, Y.~Castin, and J.~Dalibard, Phys.~Rev.~A \textbf{63},
023606 (2001). 
\bibitem{Feynman}R. P. Feynman, Int.~J.~Theor.~Phys. \textbf{21}, 467 (1982). 
\bibitem{wilson:74}K.~G.~Wilson, Phys.~Rev.~D \textbf{10}, 2445 (1974); D.~M.~Ceperley,
Rev.~Mod.~Phys. \textbf{67}, 279 (1995).
\bibitem{ceperleypimc} See e.g. D.~M.~Ceperley, \emph{Lectures on quantum monte carlo} http://archive.acsa.uiuc.edu/Science/CMP/papers/cep96b/notes.ps (1996); 
D.~M.~Ceperley, Rev.~Mod.~Phys. \textbf{67}, 279 (1995).
\bibitem{paperA}P.~Deuar and P.~D.~Drummond, J.~Phys.~A:~Math.~Gen. \textbf{39}, 1163 (2006). See also \cite{inprep}.
\bibitem{GP} For a review of Gross-Pitaevskii equations as applied to BECs, see e.g. F.~Dalfovo, S.~Giorgini, L.~P.~Pitaevskii, and S.~Stringari, Rev.~Mod.~Phys. \textbf{71}, 463 (1999).
\bibitem{evaporativecooling}P.~D.~Drummond and J.~F.~Corney, Phys.~Rev.~A \textbf{60}, R2661
(1999). 
\bibitem{samperrorinbec}M.~J.~Steel, M.~K.~Olsen, L.~I.~Plimak, P.~D.~Drummond, S.~M.~Tan,
M.~J.~Collett, D.~F.~Walls, and R.~Graham, Phys.~Rev.~A \textbf{58},
4824 (1998). 
\bibitem{inprep} In preparation. See also, P.~Deuar, PhD thesis, The University of Queensland (2004), cond-mat/0507023.
\bibitem{carterdrummond:87}S.~J.~Carter, P.~D.~Drummond, M.~D.~Reid, and R.~M.~Shelby,
Phys.~Rev.~Lett. \textbf{58}, 1841 (1987).
\bibitem{Hardman}P. D. Drummond and A. D. Hardman, Europhys. Lett. 21, 279 (1993). 
\bibitem{Solexp}M. Rosenbluh and R. M. Shelby, Phys. Rev. Lett. \textbf{66}, 153 (1991);
P.D. Drummond, R. M. Shelby, S. R. Friberg \& Y. Yamamoto, Nature
\textbf{365}, 307 (1993). 
\bibitem{carusottothermo}Y.~Castin, and I.~Carusotto, J.~Phys.~B \textbf{34}, 4589 (2001). 
\bibitem{Steel98}M.~J.~Steel, M.~K.~Olsen, L.~I.~Plimak, P.~D.~Drummond, S.~M.~Tan, M.~J.~Collett, D.~F.~Walls, and R.~Graham, Phys. Rev. A \textbf{58}, 4824
(1998).
\bibitem{Norrie}A.~A.~Norrie, R.~J.~Ballagh, and C.~W.~Gardiner, Phys.~Rev.~Lett.~\textbf{94}, 040401 (2005).

\bibitem{Kinsler} P.~Kinsler and P.~D.~Drummond, Phys. Rev. A \textbf{43}, 6194 (1991);
P.~Kinsler, M.~Fernee and P.~D.~Drummond,  Phys Rev. A \textbf{48}, 3310-3320 (1993).

\bibitem{Sinatra}A.~Sinatra, C.~Lobo, and Y.~Castin, J.~Phys.~B:~At.~Mol.~Opt.~Phys.~\textbf{35}, 3599 (2002).
\bibitem{noiseoptimisation}L.~I.~Plimak, M.~K.~Olsen, and M.~J.~Collett, Phys.~Rev.~A
\textbf{64}, 025801 (2001).
\bibitem{Timereversal}
M. R. Dowling, P. D. Drummond, M. J. Davis, and P. Deuar
Phys. Rev. Lett. \textbf{94}, 130401 (2005).
\bibitem{interactionassumptions}A treatment of these issues can be found in A.~J.~Leggett, Rev.~Mod.~Phys.
\textbf{73}, 307 (2001). For a more detailed treatment see J.~Weiner,
V.~S.~Bagnato, S.~Zilio, and P.~S.~Julienne, Rev.~Mod.~Phys.
\textbf{71}, 1 (1999). 
\bibitem{KokkelmansHolland2002}S. J. J. M. F. Kokkelmans, \emph{}\textit{\emph{J.N. Milstein, M.L.
Chiofalo, R. Walser, and M.J. Holland}}, Phys. Rev. A \textbf{65},
053617 (2002).
\bibitem{Abrikosov}A. A. Abrikosov, L. P. Gorkov, and I. E. Dzyaloshinski, \emph{Methods
of Quantum Field theory in Statistical Physics} (Dover, New York,
1963)
\bibitem{Louisell}W. H. Louisell, \emph{Quantum Statistical Properties of Radiation}
(Wiley, New York, 1973).
\bibitem{gardiner}See e.g. C.~W.~Gardiner, \textit{Quantum Noise} (Springer-Verlag,
Berlin, Heidelberg, 1991); C.~W.~Gardiner \textit{Handbook of Stochastic
Methods} (Springer-Verlag, Berlin New York, 1983). 
\bibitem{Lindblad}G. Lindblad, Comm. Math Phys. \textbf{48}, 119 (1976).
\bibitem{Greiner}M. Greiner, O. Mandel, T. W. Hänsch and I. Bloch, Nature \textbf{419},
51 (2002).
\bibitem{ccp2k}P.~Deuar and P.~D.~Drummond, Comp.~Phys.~Commun. \textbf{142},
442 (2001). 
\bibitem{healinglength}See e.g. F.~Dalfovo, S.~Giorgini, L.~P.~Pitaevskii, and S.~Stringari,
Rev.~Mod.~Phys. \textbf{71}, 463 (1999), p. 481. 
\bibitem{boundarytermerrors}A.~Gilchrist, C.~W.~Gardiner, and P.~D.~Drummond, Phys.~Rev.~A
\textbf{55}, 3014 (1997). 
\bibitem{CCconvergence}I.~Carusotto and Y.~Castin, Laser~Physics~\textbf{13}, 509 (2003).
\bibitem{2modes}See e.g. references in A.~J.~Leggett, Rev.~Mod.~Phys. \textbf{73},
307 (2001), Part VII. 
\bibitem{Donley}E. A. Donley \textit{et al}., Nature \textbf{417}, 529 (2002).
\bibitem{Bosonic-Cs-Rb-Na}J. Herbig \textit{et al}., Science \textbf{301}, 1510 (2003); K. Xu.
T. Mukaiyama \textit{et al.}, \textit{ibid.} \textbf{91} , 210402
(2003); S. D\"{u}rr, T.~Volz, A.~Marte, and G.~Rempe, Phys. Rev. Lett. \textbf{92},
020406 (2004).
\textbf{92}, 040405 (2004). 
\end{thebibliography}
\end{document}